\documentclass[preprint,noshowpacs,aps,amsfonts,showpacs,superscriptaddress,floatfix]{revtex4}
\usepackage{dcolumn}
\usepackage{color}
\usepackage{epstopdf,epsfig}
\usepackage{bm}
\usepackage{latexsym}

\newcommand\beq            {\begin{equation}}
\newcommand\bea           {\begin{equation}\begin{array}l\displaystyle}
\newcommand\eeq            {\end{equation}}
\newcommand\bes           {\begin{subequations}}
\newcommand\esu           {\end{subequations}}

\begin{document}

\title{Truncated Conformal Space Approach for Perturbed Wess-Zumino-Witten $SU(2)_k$ Models}

\author{M. Beria}
\affiliation{SISSA-International School for Advanced Studies, Via Bonomea 265,
I-34136 Trieste, Italy EU, and INFN, Sezione di Trieste.}

\author{G. P. Brandino}
\affiliation{Institute for Theoretical Physics, University of Amsterdam, Science Park 904, Postbus 94485, 1090 GL Amsterdam, The Netherlands.}

\author{Luca Lepori}
\affiliation{Departament de F\'{i}sica, Universitat Aut\`{o}noma de Barcelona, E-08193 Bellaterra, Spain.}
\affiliation{IPCMS (UMR 7504) and ISIS (UMR 7006), Universit\'{e} de Strasbourg and CNRS, Strasbourg, France.}

\author{R. M. Konik}
\affiliation{Condensed Matter and Material Science Department, Brookhaven National Laboratory, Upton NY 11973-5000, USA.}

\author{G. Sierra}
\affiliation{Instituto de Fisica Teorica, UAM-CSIC, Madrid, Spain.}

\begin{abstract}
We outline the application of the truncated conformal space approach (TCSA) to perturbations of $SU(2)_k$ Wess-Zumino-Witten theories.
As examples of this methodology, we consider two distinct perturbations of $SU(2)_1$ and one of $SU(2)_2$.
$SU(2)_1$ is first perturbed by its spin-1/2 field, 
a model which is equivalent to the sine-Gordon model at a particular
value of its coupling $\beta$.
The sine-Gordon spectrum is correctly reproduced as well as the corresponding finite size corrections.
We next  study $SU(2)_1$ with a  marginal current-current perturbation.  The TCSA results can be matched to perturbation theory
within an appropriate treatment of the UV divergences.  We find
however that these results do not match field theoretic computations
on the same model performed with a Lorentz invariant regulator.
Finally,  we consider $SU(2)_2$ 
perturbed by its spin-1
field,  which  is equivalent to three decoupled massive Majorana fermions.  
In this case as well the TCSA reproduces accurately the known 
spectrum.
\end{abstract}

\maketitle

\section{Introduction}

Correlations in one dimensional systems typically require non-perturbative, non-mean field techniques to access the underlying physics.
These techniques often trade on low energy, field theoretical  
reductions of the system.  Examples include bosonization \cite{Bosonization}, 
conformal field theory \cite{dif},
integrable field theory \cite{Muss}, the Bethe ansatz \cite{BA}, and the truncated conformal spectrum approach (TCSA) \cite{YZ}.
The  last approach, unlike the aforementioned 
techniques, is able to deal in principle with any one dimensional field theory in an exact numerical manner. 
In that sense the TCSA is similar to the density matrix renormalization group (DMRG) \cite{White}, 
but the framework where it is formulated is field theoretical
and not discrete quantum lattice systems. 

TCSA deals with  models whose Hamiltonians can be represented in the form,
\begin{equation}
H = H_{CFT} + \Phi,
\end{equation}
where $H_{CFT}$ is the Hamiltonian of some conformal field theory and $\Phi$ is an arbitrary perturbation.  The approach employs 
the Hilbert space of the conformal field theory as a computational basis and exploits the ability to compute 
matrix elements of the perturbing
field in this same basis using the constraints afforded by the conformal symmetry. 
It was first employed by Yurov and Zamolodchikov  in
studies of massive perturbations of the critical Ising model \citep{YZI} and the scaling Yang-Lee model \cite{YZ}.  

Since its introduction it has been used in a large number of instances and to some degree has become a standard tool.  It has been used
to study perturbations of the tri-critical Ising \cite{LMC,LTCM}, the 3-states Potts model 
\cite{L3PM}, bosonic $(c = 1)$ compactified theories \cite{Fev2}, the sine-Gordon model \cite{Fev3,Bajnok}, 
and perturbations of boundary 
conformal field theories \cite{Dorey3}. Spectral flows between different conformal field theories were addressed in Ref.
\cite{KlaMe2}, while the correctness of the thermodynamic Bethe ansatz
equations was checked in Refs. \cite{Dorey1,Dorey2}. Finite-size corrections to the mass spectra have also been analyzed 
in Refs. \cite{KlaMe,Fev1}. 
By replacing the conformal
field theory with an integrable field theory, the approach can also study models of perturbed integrable field theories.
Matrix elements of the perturbing field are then computed in the form factor bootstrap approach \cite{FonsecaZam}.

While the TCSA approach is extremely flexible in the models it can attack, it is in practice limited to perturbations
of conformal field theories with small central charge ($c \lesssim 1$).  For theories with large central charge,
the underlying conformal Hilbert space is large and becomes numerically burdensome to manipulate.
The difficulty has recently been partially ameliorated with the development of a numerical renormalization group (NRG)
for the TCSA \cite{NRG,NRG1,NRG2,NRG3,NRG4}.  
This renormalization group permits large Hilbert spaces to be dealt with piecewise making the numerics
manageable.  Using this renormalization group, the excitonic spectrum of semi-conducting carbon nanotubes was studied \cite{NRG3}
(here the underlying conformal field theory had $c=4$) 
as were large arrays of coupled quantum Ising chains \cite{NRG1} (here the underlying conformal field theory
had $c\sim 30-50$).

The TCSA approach, as designed, focuses on accurately computing the properties of the low energy states.  However
when combined with an NRG together with a sweeping algorithm not dissimilar to the finite volume algorithm of
the DMRG \cite{White}, the TCSA can compute the properties of states over a wide range
of energies.  This was demonstrated in \cite{NRG2} where the level spacing statistics were studied in crossing over from
an integrable to a non-integrable model.

To the best of our knowledge the TCSA has not been applied previously to perturbations of Wess-Zumino-Witten (WZW) models.  WZW
models are non-linear sigma models whose field $g$ lives on a group manifold G.  They possess topological terms, the Wess-Zumino
term, whose action is quantized with the consequence that its coupling constant $k$ is constrained to be a positive integer.  
The consequence of the topological term is to make the sigma models conformal with the 
affine Lie algebra associated with G, spectrum generating.  Affine Lie algebras typically have
richer structure than the Virasoro algebra and are consequently more difficult to treat with the TCSA.  Specialized
code needs to be developed in order to treat such models.  Here we report the development of such code for
the study of perturbations of $SU(2)_k$ WZW models.

Perturbed $SU(2)_k$ WZW models are interesting physically primarily because they are able to represent the low energy
structure of spin chains \cite{AfHal}.  The most important example here is the spin-1/2 Heisenberg chain, whose
low energy behavior is governed by $SU(2)_1$ perturbed by a marginally irrelevant current-current interaction.  Adding 
different perturbations to $SU(2)_1$ leads to different variants of the Heisenberg model.  The Hamiltonian
\begin{equation}
H = H_{SU(2)_1} + g \, \int dx \, \bar{J}_R\cdot\bar{J}_L + h\int dx \, (\phi_{1/2,1/2} \, \bar\phi_{1/2,-1/2}-\phi_{1/2,-1/2} \, \bar\phi_{1/2,1/2}) \, ,
\end{equation}
where $\phi_{1/2,\pm 1/2}$ are the spin-1/2 fields in $SU(2)_1$, is the low energy reduction of the dimerized $J_1-J_2$ Heisenberg model
\begin{equation}
H=\sum_n J_1\, (1+\delta(-1)^n) \, \bar{S}_n\cdot\bar{S}_{n+1}+J_2 \, \bar{S}_n\bar{S}_{n+1} \, .
\end{equation}
The couplings of the two models are related via $g\propto J_2-J_{2c}$ and $h \propto \delta$.
This model has been studied intensely both field theoretically with RG analyses \cite{Af1} and with DMRG \cite{kumar,branes}.
These methodologies do not currently agree on how the spin gap depends upon the dimerization parameter $\delta$.
It is one of the aims of our work to set up the framework under which disputed questions surrounding this model can be addressed.

$SU(2)_k$ for $k>1$ WZW theories are also of considerable interest as they are the low energy reductions of families of
spin-$k/2$  spin chains with finely tuned, local interactions \cite{AfHal}.  
They have also been shown more recently to represent Haldane-Shastry
type spin chains \cite{hs} with longer range interactions \cite{cs,ncs,greiter}.  In both cases, 
it is of interest to understand how $SU(2)_k$ WZW
behaves in the presence of relevant and marginal perturbations.  
And more generally, because $SU(2)_k$ WZW theories are multicritical with many possible
relevant perturbations. Actual spin chains are likely to be realized only  in the vicinity of these critical points rather than precisely
at them.

The outline of this paper is as follows.  In Section II we present briefly  the framework for the TCSA where  perturbations
of $SU(2)_k$ can be studied.  In the next three sections we  present applications of the TCSA to perturbed $SU(2)_k$ models,
establishing how the methodology works.  In Section III we examine the  $SU(2)_1 + Tr (g)$ model, which is 
 equivalent to the sine-Gordon model.
This allows us to compare our numerics with known analytic results.  In Section IV we study the  $SU(2)_1 + \bar{J}_L\cdot\bar{J}_R$
model, which corresponds to a marginal current-current perturbation of 
 $SU(2)_1$.  We show here that one can  isolate the UV divergent
behavior, essential for extracting the universal behavior of spin chains described by $SU(2)_k$ with
a marginal perturbation.  Finally, in Section V, we consider $SU(2)_2$ perturbed by $Tr (g)^2$.  This provides a useful benchmark 
of our methodology as the theory is equivalent to three non-interacting massive Majorana fermions.  We close in Section VI
with conclusions and a discussion of future directions.

\section{TCSA for $SU(2)_k$ WZW models}
\label{TCSAdesc}
In this section we describe the application of the TCSA \cite{YZ} to deformations 
of the $SU(2)_k$ Wess-Zumino-Witten model \cite{KZ,Af2}.
We thus start by considering Hamiltonians of the form
\beq
H = H_{SU(2)_k}  +   g \int^R_0 \mathrm{d} x \, \Phi (x).
\label{ham}
\eeq
Here $\Phi$ is a spin singlet combination of (highest weight or current) fields of the WZW model. 
The Hamiltonian is defined on a circle of length $R$. Adding the time coordinate
the underlying space-time is an infinite cylinder with circumference R.

The first step in the TCSA is to characterize the unperturbed theory, $H_{SU(2)_k}$.  This theory provides the computational basis
of  the TCSA numerics.
$H_{SU(2)_k}$ has central charge $c=3k/(k+2) \; (k=1,2, \dots)$,  and can be written \`{a} la  Sugawara  in terms of the  $SU(2)_k$  currents:
\begin{eqnarray}
H_{SU(2)_k} &=& \frac{2\pi}{R}\left(L_0+\bar{L}_0-\frac{c}{12}\right) \cr\cr
&=& \frac{2\pi}{R}\bigg(\sum_m :\bigg[(2J^0_mJ^0_{-m} + J^+_mJ^-_{-m} + J^-_mJ^+_{-m}) \cr
&& \hskip .6in + (2\bar{J}^0_m\bar{J}^0_{-m} 
+ \bar{J}^+_m\bar{J}^-_{-m} + \bar{J}^-_m\bar{J}^+_{-m})\bigg]: - \frac{c}{12}\bigg). 
\end{eqnarray}
The left-moving currents $J^{0,\pm}_m$ obey the algebra,
\begin{eqnarray}\label{current-algebra}
& &\left[J_{m}^{0},J_{n}^{0}\right]=\frac{km}{2} \, \delta_{n+m,0}, \nonumber\\
& &\left[J_{m}^{0},J_{n}^{\pm}\right]=\pm J_{m+n}^{\pm}, \\
& &\left[J_{m}^{+},J_{n}^{-}\right]=2J_{m+n}^{0}+km \, \delta_{m+n,0},  \nonumber \, 
\end{eqnarray}
with the right moving currents $\bar{J}^{0,\pm}_m$  obeying the same  algebra. 

The field content of $H_{SU(2)_k}$ consists of $k+1$ primary fields, $\phi_{s,m=-s,\cdots,s}$, forming spin $s=0,\cdots,k/2$ representations.
The conformal weight of the spin $s$ primary field is given by $\Delta_s = s(s+1)/(k+2)$. 
The primary  fields and the WZW currents are the basic tools to construct the Hilbert space of the unperturbed theory, providing the first  ingredient of  the TCSA.
The Hilbert space can be written as a tensor product of its holomorphic (left) and anti-holomorphic (right)  degrees of freedom.  
For $SU(2)_k$, the left and right  sectors of the Hilbert space each have $k+1$ modules, where  each module is associated to one primary field.
The module consists of a highest weight state, $|s,s\rangle \equiv \phi_{s,s}(0)|0\rangle$, together with an infinite tower of 
descendant states:
\begin{equation}
J_{-n_M}^{a_M}\ldots J_{-n_1}^{a_1}|\phi_{s,s}\rangle, ~~~ n_i = 0,1,2,\cdots ~~~ a_i=0,\pm.
\label{ostates}
\end{equation}
These states (\ref{ostates}) are eigenstates of the Virasoro operator $L_{0}$,  

and the third component component of the current  $J^{0}_{0}$:
\begin{eqnarray}
& &L_{0}\left(J_{-n_{M}}^{a_{M}}\ldots J_{-n_{1}}^{a_{1}}\right)|s,s\rangle=\left[\Delta_{s}+\sum_{i}n_{i}\right]
\left(J_{-n_{M}}^{a_{M}}\ldots J_{-n_{1}}^{a_{1}}\right)|s,s\rangle;\nonumber\\
&&{}\\
& &J^{0}_{0}\left(J_{-n_{M}}^{a_{M}}\ldots J_{-n_{1}}^{a_{1}}\right)|s,s\rangle
=\left[s+\sum_{i}a_{i}\right]\left(J_{-n_{M}}^{a_{M}}\ldots J_{-n_{1}}^{a_{1}}\right)|s,s\rangle\nonumber .
\end{eqnarray}
The quantity $\sum_{i} n_i$ is called the Kac-Moody level, or simply the level,  of the descendant  state.

The set of states (\ref{ostates}) is not yet a basis of the Hilbert space  since  it is over complete and contains null states.
In order to form a complete orthonormal basis, we tackle each Kac-Moody module separately.

We do so in an iterative fashion.  At each step we  have a set of non-zero norm states which  are linearly  independent 
(at the beginning  this set will consist solely of the highest weight state).
We next add a new descendant state to this set, compute the matrix of scalar products
of states in the expanded set (the Gramm matrix), and find its determinant. If it is non-zero, the new
state is added to the list; otherwise it is discarded.  We then move to the next descendant in the tower of states in
increasing order in the level. The process ends when all the states up to a given level have been considered. 

This procedure yields a complete set of states that
we can easily orthonormalize to obtain a basis. 
In order to optimize this procedure we take into account the following properties:
\begin{itemize} 
\item states with different $L_0$ quantum number are independent;
\item states having different spin (eigenvalue of $J_0^0$) are independent;
\item we discard the states with null norm;
\item we act only with level 0 currents, $J^{0,\pm}_0$, directly on the highest weight state $|s,s\rangle$ in the module. 
\end{itemize}
To demonstrate that the above method amounts to a numerically intensive task, we present in Table  I
the number of states per level for the two modules of $SU(2)_1$
and the three modules of $SU(2)_2$.
\begin{table}\label{chiral}
    \begin{tabular}{|c||c|c||c|c|c|}
        \hline
          & $SU(2)_1$ &  &    $SU(2)_2$   & &  \\ \hline
        level  &\textbf{I}  & \textbf{$1/2$} &  \textbf{I} & \textbf{$1/2$} & \textbf{$1$} \\ \hline
      \hline
      	\textbf{0} 	& 1	& 2	&	 	 1 	& 2	& 3	 \\ \hline 
		\textbf{1} &	 4	& 4	&		 	4 & 	8 & 7	 \\ \hline 
		\textbf{2} &	 8	& 10	&	 13 	& 	20 & 	19 \\ \hline 
		\textbf{3} &	 15	& 18	& 	 28 	& 	46 & 	40 \\ \hline 
		\textbf{4} &	 28	& 32	& 	 	58 & 	94 & 	83 \\ \hline 
		\textbf{5} &	 47	& 52	& 	 	112 & 	178 & 	 152 \\ \hline 
		\textbf{6} &	 76	& 86	& 	 	206 & 	324 & 	 275 \\ \hline 
		\textbf{7} &	 119	& 132			& 359 	& 564 	& 468	 \\ \hline 
		\textbf{8} &	 181	& 202			& 	611 & 	948 & 	 786 \\ \hline 
		\textbf{9} &	 271	& 298			& 	1002 & 	1552 & 	1272 \\ \hline 
		\textbf{10} &	 397	& 436			& 	1611 & 	2482 & 	 2026\\ \hline 
		\textbf{11} &	 571	& 622			& 2529	& 	3886 & 	 3145\\ \hline 
    \end{tabular}
    \caption{The dimensions of the Verma modules of $SU(2)_1$ and $SU(2)_2$ at different levels. }
\end{table}

Once  the chiral sector of $SU(2)_k$ has been obtained, the total Hilbert space is constructed as a tensor product of the isomorphic holomorphic
and antiholomorphic sectors.
These tensor products are diagonal in the modules, i.e. left moving spin $s$  states are only tensored with their right moving spin $s$  counterparts.
In forming these tensor products, we group the states by their value of Lorentz spin, $(L_0 - \bar{L}_0)$, and $z$-component of $SU(2)$ spin, $J^0_0 + \bar{J}^0_0$.
Recall that the Lorentz spin is proportional to the momentum carried by the corresponding state. 
In Table  \ref{nonchiral} we present the number of cumulative states up to  a given level 
in $SU(2)_1$ and $SU(2)_2$ with vanishing  Lorentz spin. 

\begin{table}\label{nonchiral}
    \begin{tabular}{|c|c|c||c|c|c|}
        \hline
      $SU(2)_1$  level    & mom. $0$  & $SU(2)$ spin 0  &  $SU(2)_2$ level &  mom. $0$    &  $SU(2)$ spin 0  \\ \hline
  			1	  &     18    &   8     &  1     &  75  & 25 \\ \hline
  			2  &  70 & 24 &  2 & 444  & 120 \\ \hline
  			5  & 1309 & 381 & 4& 6839 & 1595\\ \hline
  			11 & 133123 &  32021 & 7 & 185111 & 38665\\ \hline
    \end{tabular}
    \caption{The cumulative dimensions for non chiral spaces at given level with total Lorentz spin (mom.) zero (first column) 
and both Lorentz spin and $SU(2)$ spin $S_z$ zero (second column).}
\end{table}

Once the computational basis has been  constructed, the TCSA requires the evaluation  of matrix elements of the perturbing operator in that basis.  
For this purpose one uses  the
commutation relations of the fields, $\phi_{s,m}(0)$, with the current modes, $J^{0,\pm}_0$, which are given by
\begin{eqnarray}
& &\left[J^{0}_{0},\phi_{s,m}\right]=m \, \phi_{s,m};\nonumber\\
& &\left[J^{\pm}_{0},\phi_{s,m}\right]=(s\pm  m) \, \phi_{s,m\pm1}.
\end{eqnarray}
With these commutation relations and those of the current modes (\ref{current-algebra}), matrix elements of the
form
\begin{equation}
\langle \phi_{s,s}|J^{a_1}_{n_1}\cdots J^{a_m}_{n_m}\phi_{s',m}(0,0)J^{b_1}_{-l_1}\cdots J^{b_k}_{-l_k}|\phi_{s'',s''}\rangle,
\end{equation}
can be reduced to the structure constants, $C_{s_1,m_1,;s_2,m_2;s_3,m_3}$:
\begin{equation}
\langle \phi_{s,s}|\phi_{s'm}(0)|\phi_{s'',s''}\rangle =\left (\frac{2\pi}{R}\right)^{2\Delta_s}C_{s,s;s',m;s'',s''}.
\end{equation}
We list all non-zero structure constants, $C_{s_1,m_1,;s_2,m_2;s_3,m_3}$ in Table III for $SU(2)_1$ and $SU(2)_2$.

\begin{table}\label{tab_struc.constants}
    \begin{tabular}{|cc|cc|}
        \hline
         $SU(2)_1$ &  & &  \\ \hline
        $C_{0,0;1/2,-1/2;1/2,1/2}$ & 1 & $C_{1/2,1/2;1/2,1/2;0,0}$ & 1 \\ \hline  \hline
     $SU(2)_2$ &  & &  \\ \hline
        $C_{0,0;1/2,-1/2;1/2,1/2}$ & 1 & $C_{1,1;1,1;0,0}$ & 1 \\ 
        $C_{0,0;1,-1;1,1}$ & 1 & $C_{1,1;1/2,1/2;1/2,1/2}$ &1\\ 
     $C_{1/2,1/2;1/2,1/2;0,0}$ & 1 & $C_{1/2,1/2;1/2,-1/2;1,1}$ & -1\\ 
  $C_{1/2,1/2;1,0;1/2,1/2}$ & $-\frac{1}{\sqrt{2}}$ &  & \\ \hline
    \end{tabular}
\caption{Non zero structure constants of $SU(2)_1$ and $SU(2)_2$.}
\end{table}

With the spectrum of the unperturbed $SU(2)_k$ model specified and the matrix elements of the perturbing field given,
we are able to compute explicitly the matrix elements of the full Hamiltonian on the circle  and represent it in
matrix form.  To be able to analyze this Hamiltonian, we truncate (the truncation in the acronym TCSA) 
the Hilbert space by discarding all states with a chiral component whose level is greater than $N_{tr}$. 
The resulting \emph{finite} dimensional Hamiltonian matrix can then be diagonalized numerically, obtaining the spectrum of 
the perturbed  theory.  This procedure is particularly robust for relevant perturbations  since  the low energy
eigenstates of the perturbed theory, $|r \rangle$, are localized on the low energy conformal states, $\{ |c\rangle_\alpha \}$ . Namely,  
expanding  $|r \rangle$ into the  conformal basis, $\{ |c\rangle_\alpha \}$,
\begin{equation}
|r\rangle = \sum_\alpha b_\alpha |c\rangle_\alpha,
\end{equation}
 the coefficients $b_\alpha$, are primarily concentrated  on the  low energy 
conformal states $|c\rangle_\alpha$ determined by $H_{SU(2)_k}$. 

To extract  physical quantities  for  the perturbed Hamiltonian (\ref{ham}), such as
the mass gap, energy levels, correlation functions, etc., 
the choice of the system size, $R$, requires special consideration. 
For $mR \ll 1$ (here $m$ is the putative mass scale of the perturbed theory), the system lies in the UV
limit where  the conformal term, $H_{SU(2)_k}$, of the full Hamiltonian dominates.  In this regime the spectrum
resembles that of the conformal $H_{SU(2)_k}$ where the  energy levels scales as $1/R$.
In the IR regime, $Rm \gg 1$, the perturbation, $\int \!\!dx \Phi(x)$, dominates and one expects a scaling of the form 
$\sim R^{1-2 \Delta_s }$, where $2 \Delta_s$ is the scaling dimension of the perturbing field.
In general,  the spectrum of the perturbed model  must be extracted in a region of $R$ where the conformal term 
and the perturbation are balanced in the sense that physical quantities
remains stable under small variations of $R$. This region is usually denoted  as the ``physical window''. 

For theories where the dimension of the Hilbert space grows very fast, i.e. $SU(2)_k$ with $k$  large,
the truncation scheme proposed above may  not yield accurate results. 

In those cases one can take recourse to
a numerical renormalization group (NRG) improvement of the TCSA \cite{NRG,NRG1,NRG2,NRG3,NRG4}.  This procedure allows the TCSA
to reach much higher truncation levels than that possible in its unadorned form.  Taken together with
an analytic renormalization group, it is possible to remove the effects of truncation altogether \cite{NRG}.
While this NRG has been tested extensively on relevant perturbations of conformal field theories,
it has not hitherto been tried on marginal perturbations of CFTs.  We will show in Section IV that the NRG can accurately
predict the low lying spectrum even in the marginal case.

\section{$SU(2)_1$ perturbed by the spin-1/2 field}
\label{SGsection}
Our first test of the TCSA is  the perturbation of the $SU(2)_1$ WZW model by the 
singlet formed from the spin-1/2 operator of $SU(2)_1$, $\Phi  = \left(\phi_{1/2,1/2} \, \bar\phi_{1/2,-1/2}-\phi_{1/2,-1/2} \, \bar\phi_{1/2,1/2}\right)$: 
\beq
H_{rel} = H_{SU(2)_1}  +  h \int \mathrm{d} x \,  \Phi(x) \, .
\label{hamrel}
\eeq
The scaling dimension of $\Phi$ is $ 2 \Delta_{1/2} = 1/2$, and so this is a relevant perturbation. 
Moreover  the theory is invariant under $h \to -h$. 
This Hamiltonian is equivalent to the sine-Gordon model whose Lagrangian is given by 
 \cite{tsvbook}
\beq
L_{SG} = \int \mathrm{d}^2 x \, \Big( \frac{1}{2} (\partial_{\mu} \varphi)^ 2 + 2 \lambda \cos(\beta \varphi) \Big),
\label{SG}
\eeq 
with $\beta^2 = 2 \pi$.  The  correspondence between (\ref{hamrel}) and (\ref{SG}) is based on the identifications
\begin{equation}
\phi_{1/2,1/2}\, (z) =  \mathrm{exp} \left(i \frac{\varphi(z)}{\sqrt{2}}\right) ; ~~~~\phi_{1/2,-1/2}\, (z) =  \mathrm{exp} \left(-i \frac{\varphi(z)}{\sqrt{2}}\right),
\end{equation}
which, up to phase redefinitions, are the bosonization formulas of the $SU(2)_1$ model \cite{dif}.

We now demonstrate that the TCSA numerics reproduce the expected behavior of the sine-Gordon model at this value of the SG  coupling.
The spectrum of the sine-Gordon model at $\beta^2 = 2 \pi$ is composed of a soliton $S$ and antisoliton $\bar{S}$ with mass $M$ 
and two breathers $B_1$ and $B_2$ with masses $M_1=M$ and $M_2=\sqrt{3} M $ respectively.  The soliton, anti-soliton, and the first
breather form a triplet under $SU(2)$.  The charges of the particles ($S,B_1,\bar{S}$) are given by (1,0,-1) 
and they coincide with  their $S_z$ quantum number.  The second breather, $B_2$, is  a singlet under $SU(2)$.

\begin{figure}
\centering
\includegraphics[scale=0.4]{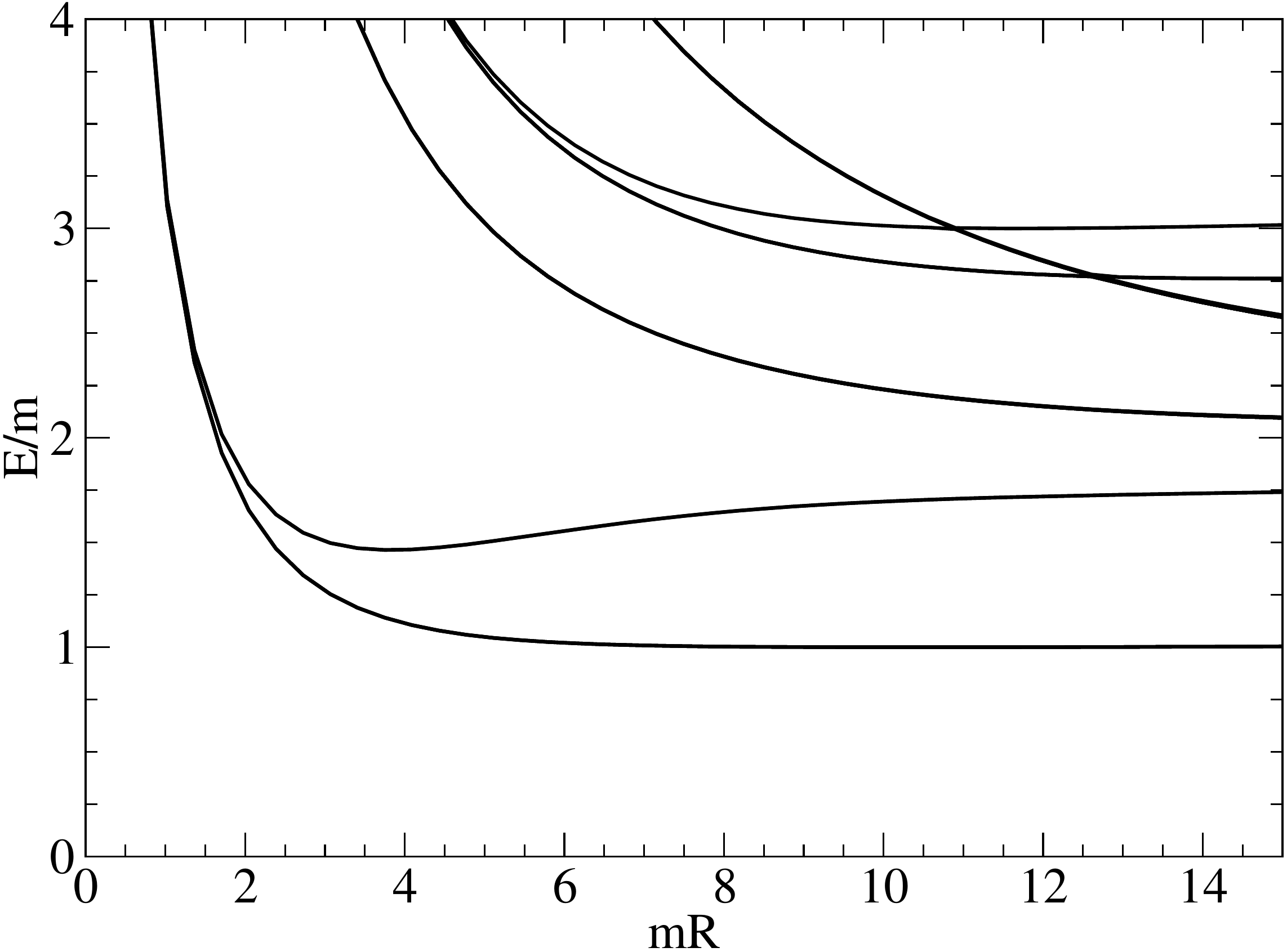}
\caption{Plot of  the TCSA data for the lowest  six excited states in the $S_z=0$ sector
with the ground state energy subtracted.  The lowest  two excited states correspond to $B_1$ (the $S_z=0$ state
of the triplet) and $B_2$.  The next four excited states are two particle states.
The data are  computed with a truncation level $N_{tr}=9$.
The fundamental triplet of particles can be seen to have mass M=1.0016. } 
\label{SGspectrum}
\end{figure}

Fig. \ref{SGspectrum} shows the low energy TCSA spectrum which reproduces the  basic structure: a low lying  triplet
with mass $M$,  and a single excitation at roughly $\sqrt{3}M$.  The expected value of the mass $M$ can
be determined from the coupling constant used in the TCSA.
The relation between the coupling of the sine-Gordon model and the mass M is given as \cite{SGmass}
\begin{equation}
h=\lambda=\frac{\Gamma(\frac{\beta^2}{8\pi})}{\pi  \Gamma(1-\frac{\beta^2}{8\pi})}\left[M\sqrt{\pi} \,
\frac{\Gamma\left(\frac{1}{2}+\frac{\xi}{2\pi}\right)}{2\Gamma\left(\frac{\xi}{2\pi}\right)} \right]^{2-\frac{\beta^2}{4\pi}},
\end{equation}
where $\xi=\frac{\beta^2}{8}\frac{1}{1-\frac{\beta^2}{8\pi}}=\frac{\pi}{3}$.
With $h=0.0942753..$ we expect the mass to be $1$ while from the TCSA we find $M=1.0016$, in excellent agreement.

\begin{figure} 
 \centering
\includegraphics[scale=0.4]{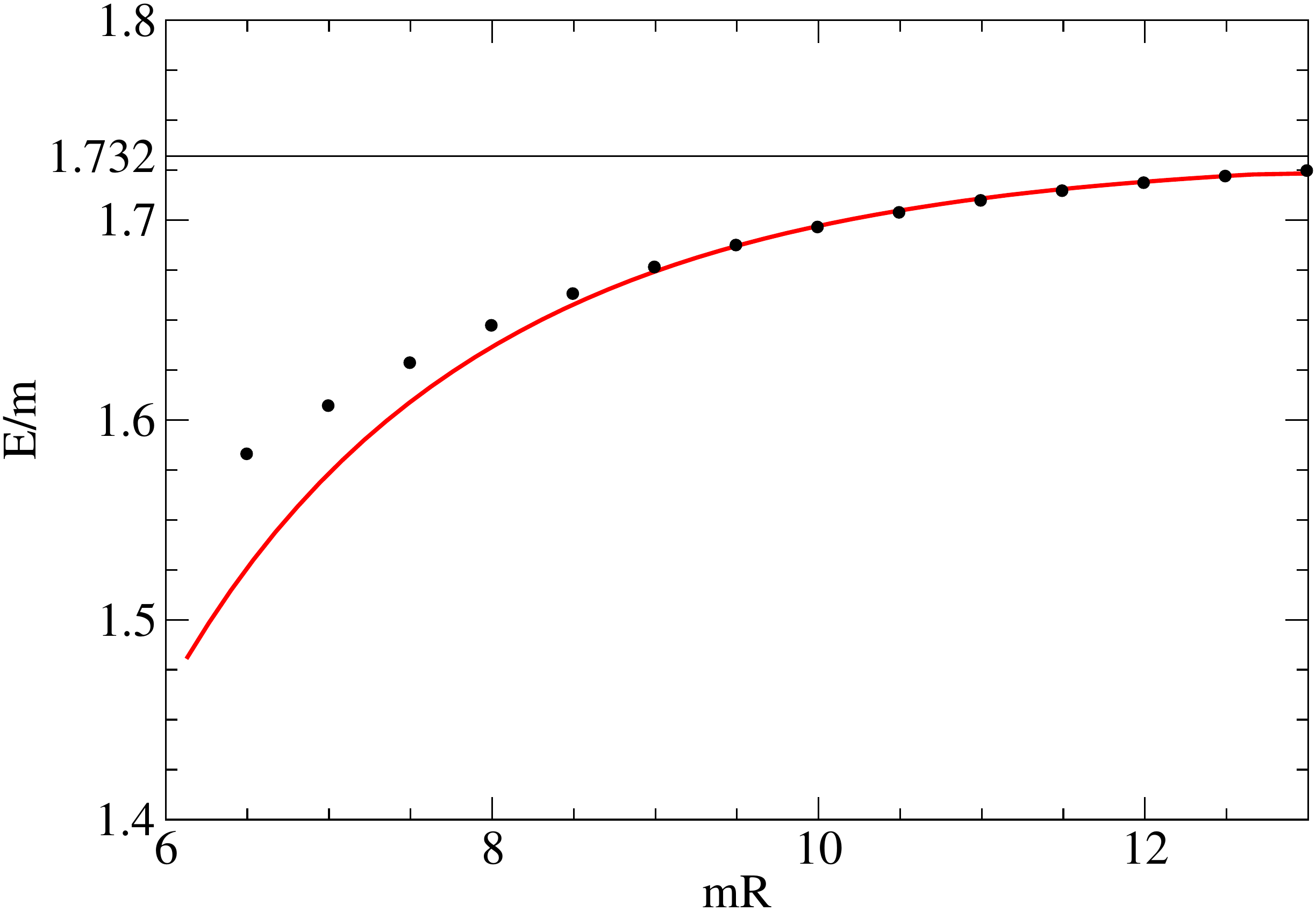}
\caption{Single particle state $B_2$: Analytic prediction for its mass with finite size effects (continuous line) compared to the TCSA data.}
\label{SingleB2}
\end{figure}

For the second breather, found roughly  at $\sqrt{3}M$, a more careful analysis of the TCSA data is required.  
For this excitation there are significant finite size corrections.  These corrections can
be understood as virtual processes on the cylinder \cite{KlaMe,YZ},  which are suppressed exponentially
as $R$ becomes large.
For $B_2$  these corrections are  given by  \cite{KlaMe}
\begin{eqnarray}
\Delta m_{B_2} (R) & =& - 3\sqrt{3} e^{-R/2} \cr\cr
& -&\int^{\infty}_{- \infty} \frac{\mathrm{d} \theta}{2 \pi} e^{-m_{B_2} R \, \mathrm{cosh} \theta} m_{B_2} R\cosh(\theta ) 
\big( S_{B_2B_2}^{B_2B_2} (\theta + i \pi/2) - 1 \big) + \mathcal O(e^{- \sigma_{B_2} R}).
\end{eqnarray}
We can estimate the so-called error exponent to be $\sigma_{B_2}=1.105$ \cite{KlaMe}.
Here $S_{B_2B_2}^{B_2B_2}(\theta)$ is the scattering $S$-matrix of the process $B_2+B_2\rightarrow B_2+B_2$ \cite{2Zam}.
Fitting  $m_{B_2}+\Delta m_{B_2}(R)$ to the TCSA data,  we find that $M_{B_2}/M=1.7322\pm 0.0003$. 
Using this mass, we plot $m_{B_2}+\Delta m_{B_2}$ against the TCSA data in Fig. \ref{SingleB2}, which 
shows  an excellent agreement  between the theory and the TCSA data for $R>9$.

\begin{figure} 
 \centering
\includegraphics[scale=0.4]{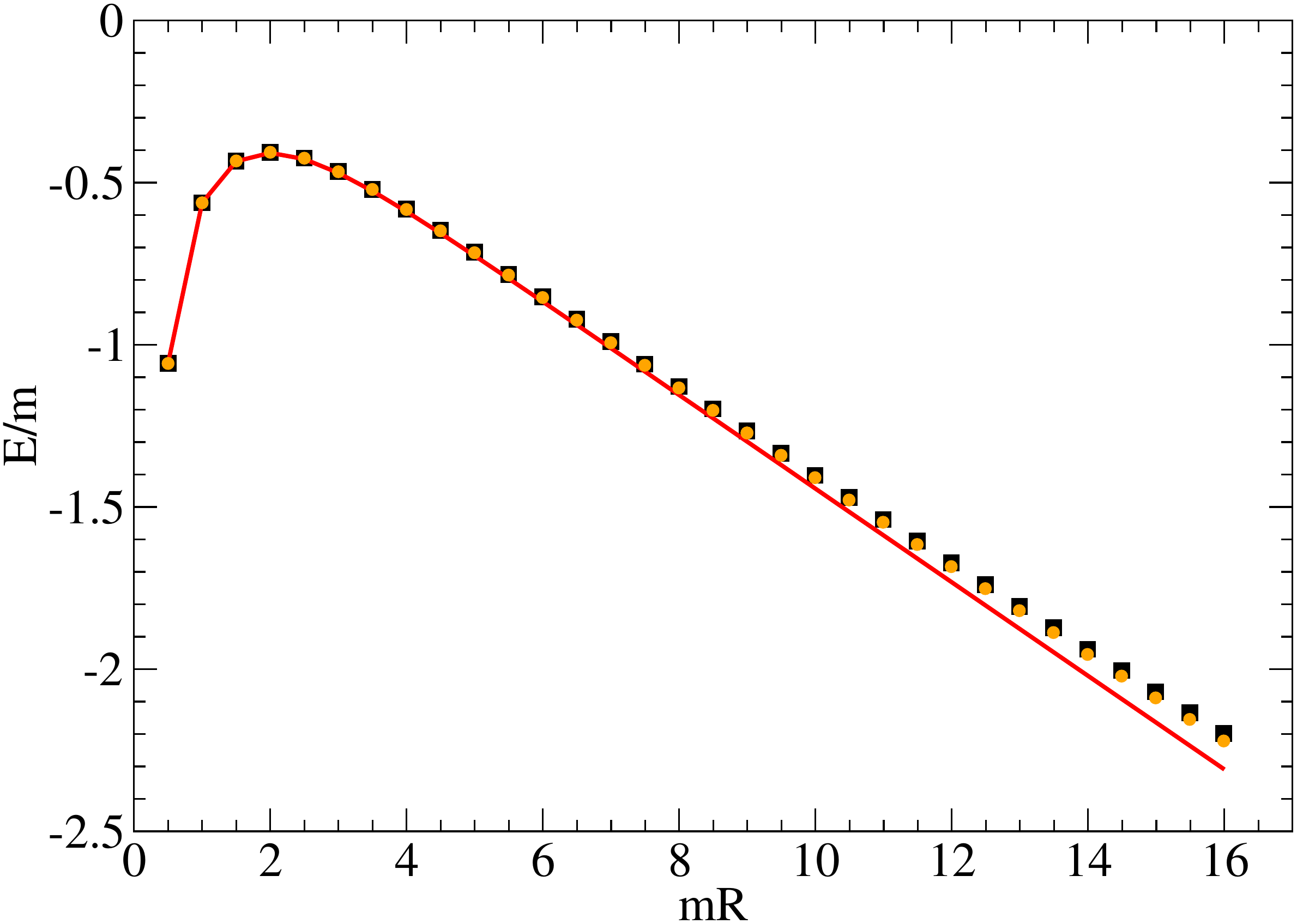}
\caption{Ground state energy: analytic prediction (continuous line) compared to the TCSA results. 
Black squares: TCSA with $N_{tr}=9$; orange circles: TCSA with $N_{tr}=10$.}

\label{TBAgs}
\end{figure}

We now turn to the ground state energy $E_{gs}$.  The TCSA gives $E_{gs}$ as would be computed in
conformal perturbation theory to all orders.  This perturbative energy can be expressed
as the sum of a linear term in $R$, proportional to a bulk energy density, $\epsilon_{\rm bulk}$, plus
a term given by the thermodynamic Bethe ansatz $E_{TBA}$ \cite{Zam3StatePotts,YZ}.

\begin{equation}
E_{gs} =  \epsilon_{\rm bulk} R  + E_{TBA}(R).
\end{equation}
The bulk contribution to $E_{gs}$ is given by \cite{SGfreeEn}
\begin{equation}
\epsilon_{\rm bulk}= -\frac{M^2}{4}\tan{\frac{\xi}{2}}=-\alpha M^2R.
\end{equation} 
For sine-Gordon with $\beta^2=2\pi$, $\alpha= 0.14438...$.  The contribution from $E_{TBA}(R)$ is given by the
solution of a coupled set of integral equations  involving the $S$-matrices of the
various excitations in the model \cite{Zam3StatePotts,YZ}.  At large $R$, this contribution reads
\begin{eqnarray}
E_{TBA}(R) &=& -3M\int^\infty_{-\infty}\frac{d\theta}{2\pi}\cosh (\theta) \, e^{-MR\cosh (\theta)}\cr\cr
&&-\sqrt{3}M\int^\infty_{-\infty}\frac{d\theta}{2\pi}\cosh (\theta) \, e^{-\sqrt{3}MR\cosh (\theta)}+{\cal O}(e^{-2MR}),
\end{eqnarray}
and essentially marks the correction to  the energy due to the spontaneous emission of a virtual particle from the vacuum
which travels around the system before being reabsorbed.
Fig. \ref{TBAgs} shows  the TCSA data against  the theoretical values of  $E_{gs}$ (including the full, not just the leading order
large R, contribution coming from $E_{TBA}(R))$. 
There is  a  good agreement for R smaller than $mR=8$ and then slight deviations thereafter, which
can be reduced by increasing the value of $N_{tr}$ (see Fig. \ref{TBAgs}). 

Above the single particle states, one  encounters  sets of two-particle states consisting of pairs
of particles from the triplet and the singlet of the SG model. 
These two-particle states can  be organized into $SU(2)$  multiplets.  For example, two-particle states
involving the triplet decompose as 
$({\bf 3} \otimes {\bf 3}) = ({\bf 5} \oplus {\bf 3} \oplus {\bf 1})$. 
These states suffer finite size corrections  due to scattering between the particles.  These effects can be taken into 
account by solving the quantization conditions for the momentum in finite volume: 
\begin{equation}\label{quant}
2\pi n_i  = m R \sinh \theta_i -i\ln{S_{ij}(\theta_i-\theta_j)}, ~~~ i=1,2, 
\end{equation} 
where  $\theta_i \;  \; (i=1,2)$ are the rapidities of the particles that parametrize their 
energy-momentum, $(E,p) = \left(m\cosh \theta_i , m\sinh \theta_i \right)$, and   
$S_{ij}$ is the scattering matrix between the two particles  \cite{Z2}.
\begin{figure}
 \centering
\includegraphics[scale=0.25]{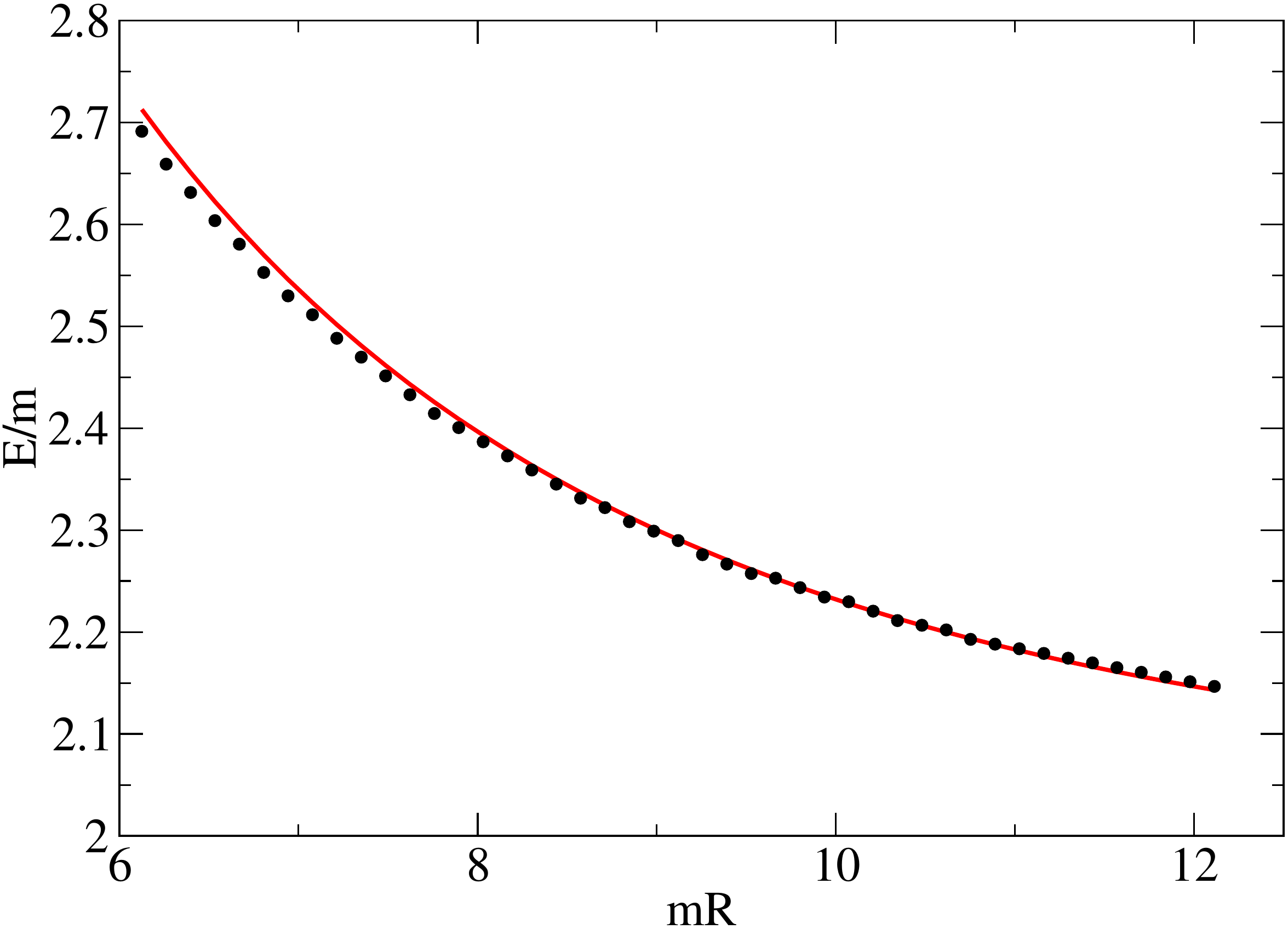}\includegraphics[scale=0.25]{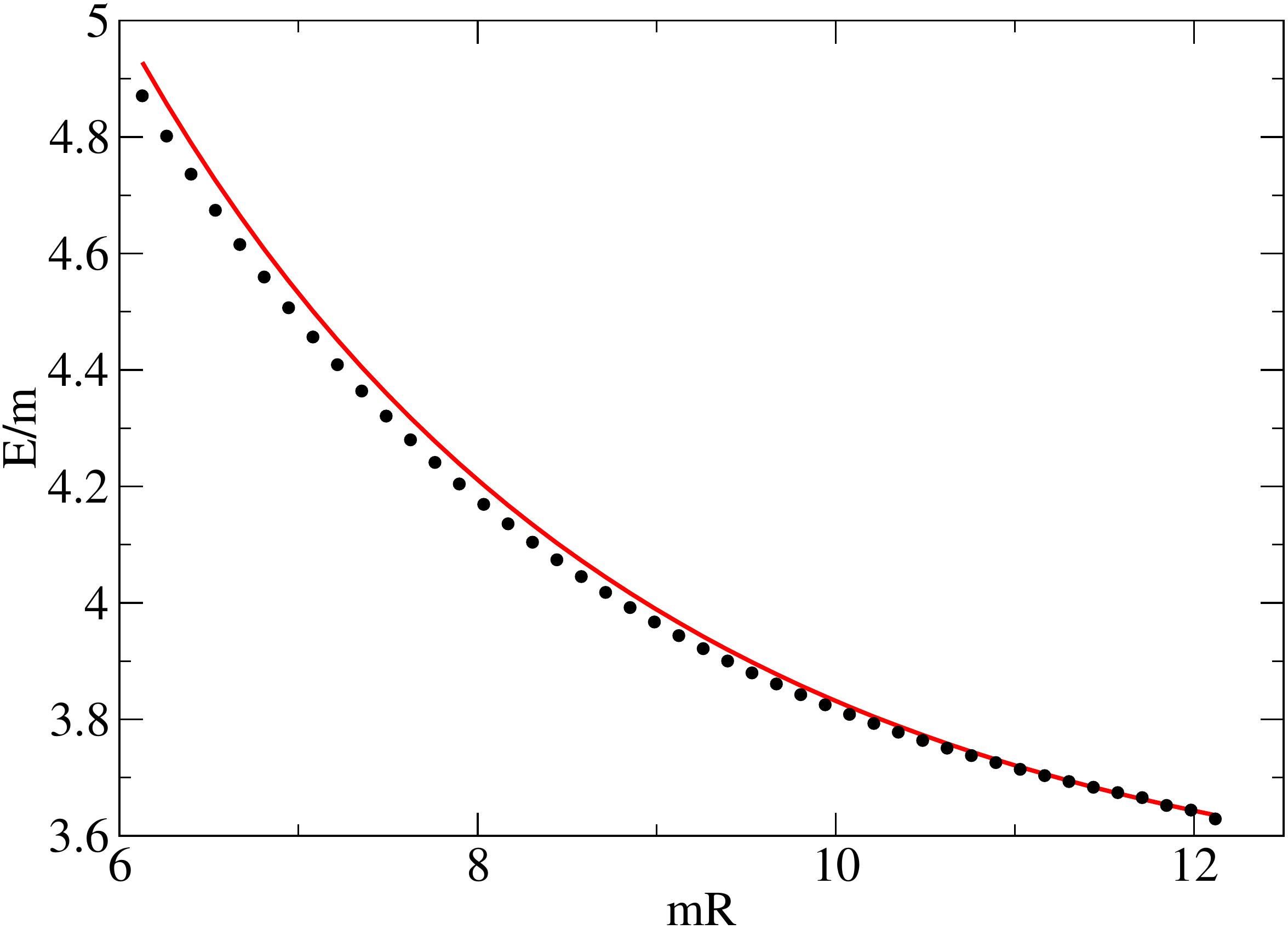}
\caption{Two-particle states: comparison between the analytic results and the TCSA data.
Left panel: a two particle state involving two triplet particles with total $S_z=0$ and $(n_1,n_2)=(-1,1)$. 
Right panel: a two-breather $B_2$ state with $(n_1,n_2)=(-1,1)$. 
Continuous line: analytical results derived from Eqn. (\ref{quant}). Dots: TCSA data.}
\label{BA}
\end{figure}
The solution of  these equations, for a pair  $(n_1,n_2)$,  yields  the rapidities  as a function of $R$, and so 
the energy of these  states.  Fig. \ref{BA} shows reasonably good agreement between the
analytical and the TCSA results, particularly at large values of $R$.  At smaller $R$,
single particle virtual processes become important, leading to deviations between the TCSA and
our analytic estimates.

\section{$SU(2)_1$ perturbed by current-current interactions}
\label{solomarginal}
In this section we consider the perturbation of the $SU(2)_1$ WZW model by the marginal current-current operator,
\beq
H = H_{SU(2)_1} -  g  \int \mathrm{d} x  \, \bar{J}_L(x)\cdot\bar{J}_R(x).
\label{hamrel2}
\eeq
Unlike the perturbation by the spin-1/2 field, here the sign of $g$ matters   
as it differentiates the model's behavior in the IR  limit \cite{2Zam}.
For $g > 0$ the perturbation is marginally relevant and asymptotically free.  The corresponding theory is 
a massive integrable rational field theory (RFT) coinciding with the $SU(2)$ Thirring model \cite{Thir}.
And for $g < 0$  the coupling is marginally irrelevant and the theory undergoes a massless RG flow 
towards the $SU(2)_1$ fixed point. 

We address two questions here.  We first ask if we can determine with the TCSA the universal correction (as explained below) to the ground state
energy due to the marginal perturbation.  
And second, we question if the numerical renormalization group improvement of the TCSA 
works in the context of marginal perturbations \cite{NRG,NRG1,NRG2,NRG3,NRG4}.

\subsection{Universal Term in Ground State Energy from the TCSA}

\begin{figure}[t]
\centering
\includegraphics[scale=0.4]{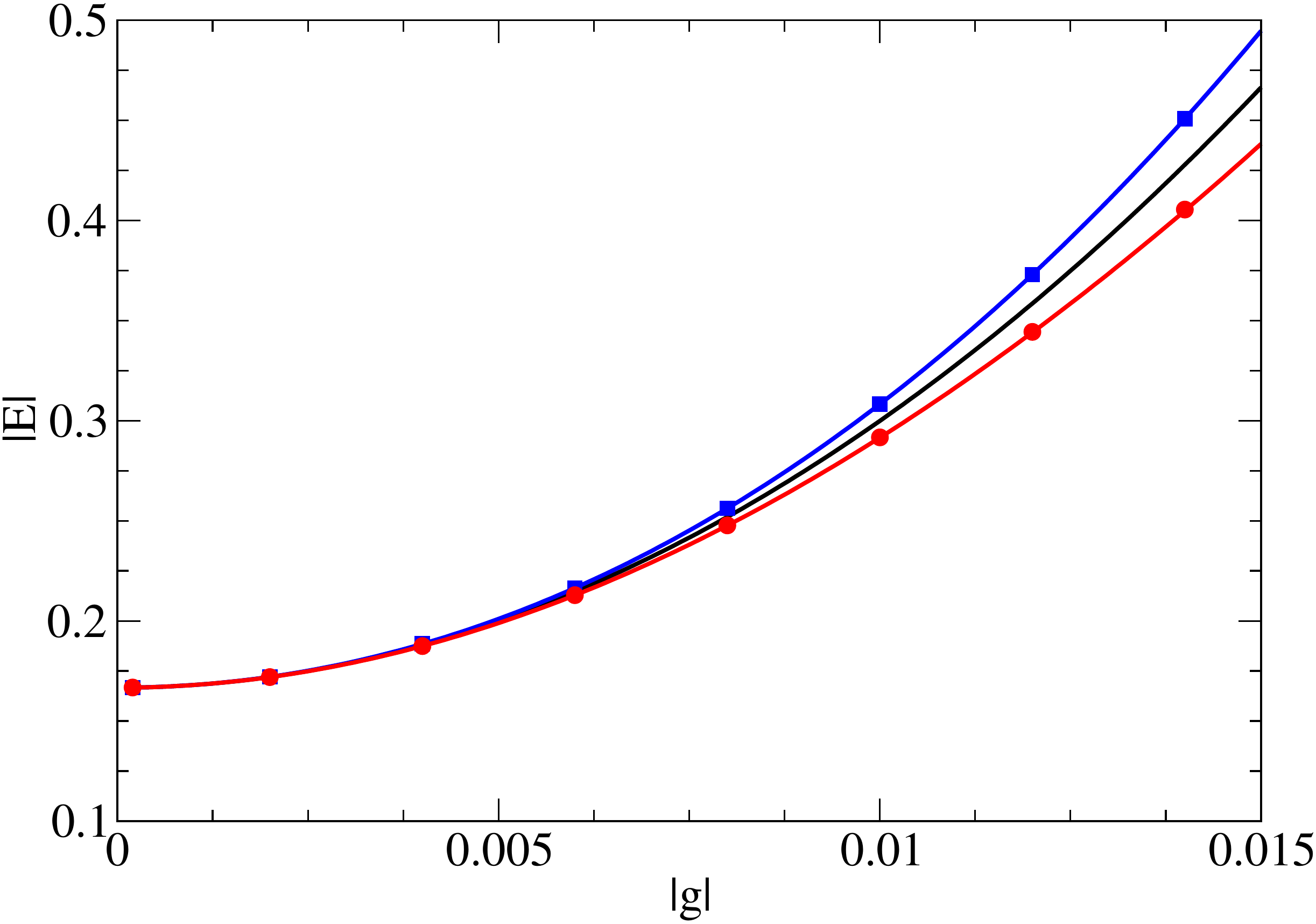}
\caption{Plots of the ground state energy of $SU(2)_1+\bar{J}_L\cdot\bar{J}_R$ as a function of the marginal coupling $g$.  Solid lines give
the perturbative computation for both $g>0$ relevant (blue) and $g<0$ irrelevant (red).  
The black line shows the second order perturbative correction in $g$. The data points represent the corresponding
numerical data from the TCSA.}
\label{marginal}
\end{figure}

In the small $g$ regime, the correction to the ground state energy on the cylinder
is given in  perturbation theory by  \cite{Cardy}:
\begin{eqnarray}
E_0 &=& -\frac{\pi}{6R}c-\frac{g^2}{2!}R\left(\frac{2\pi}{R}\right)^{2x-2}\frac{3}{4}I_2 -\frac{g^3}{3!}bR\left(\frac{2\pi}{R}\right)^{3x-4} I_3,
\end{eqnarray}
where
\begin{eqnarray}\label{integrals}
I_2 &=& \int d^2z |z|^{x-2}|z-1|^{-2x};\cr\cr
I_3 &=& \int d^2z_1 d^2z_2 |z_1|^{x-2}|z_2|^{x-2}|z_1-1|^{-x}|z_2-1|^{-x}|z_1-z_2|^{-x}.
\label{scalAA}
\end{eqnarray}
We write these expressions so that they are valid for a general
dimension, $x$, of the perturbing operator,
$\phi(=\bar{J}_L\cdot\bar{J}_R$ for $x=2$).  Our conventions are such
that two point function of $\phi$ is (on the plane)
$$
\langle \phi(r)\phi(0)\rangle = \frac{3}{4}\frac{1}{|r|^{2x}},
$$
while its corresponding three point function equals
\begin{equation}
\langle \phi(r_1)\phi(r_2)\phi (r_3) \rangle = \frac{b}{|r_{12}|^x|r_{13}|^x|r_{23}|^x},
\end{equation}
with $b=3/2$.   The $\beta-$function for this theory in these
conventions is given by 
\begin{equation}
\frac{d \tilde g}{dl} = (2-x)\tilde g + \frac{4\pi
  b}{3\Gamma^2(x/2)}\tilde g^2.
\end{equation}
where $\tilde g$ is then the dimensionless coupling and $l$ is a
logarithmic length scale.
To convert to the conventions of
Refs. \cite{Cardy,LudwigCardy}, we need to take $\phi \rightarrow -\sqrt{\frac{4}{3}}\phi$.

Although the theory is defined on a spacetime cylinder, under a conformal transformation the integrals can be written as
integrals over the plane, as given  above.
The integrals $I_2$ and $I_3$ are UV divergent and require regulation.  The divergent pieces of these integrals contribute to
the non-universal piece of $E_0$ (non-universal because their value depends on the regulation scheme).  
If $\epsilon_p$ is a short distance regulator in the plane (i.e. the difference of two plane integration variables is not permitted to 
be smaller than $\epsilon_p$), both $I_2$ and $I_3$ contain terms
proportional to $\epsilon_p^{-2}$ (for $x=2$).  Under a conformal transformation,
$\epsilon_p$ is related to a short distance regulator on the cylinder, $\epsilon_c$, by $\epsilon_p=2\pi\epsilon_c/R$.  Thus the divergent
pieces lead non-universal correction of the form
\begin{equation}
E_{0,non-univ} = (a_2g^2 + a_3g^3 + \cdots) \, \frac{R}{\epsilon_c^{2x-2}},
\end{equation}
i.e. these corrections scale linearly with the system size.
\begin{figure}[t]
\centering
\includegraphics[scale=0.4]{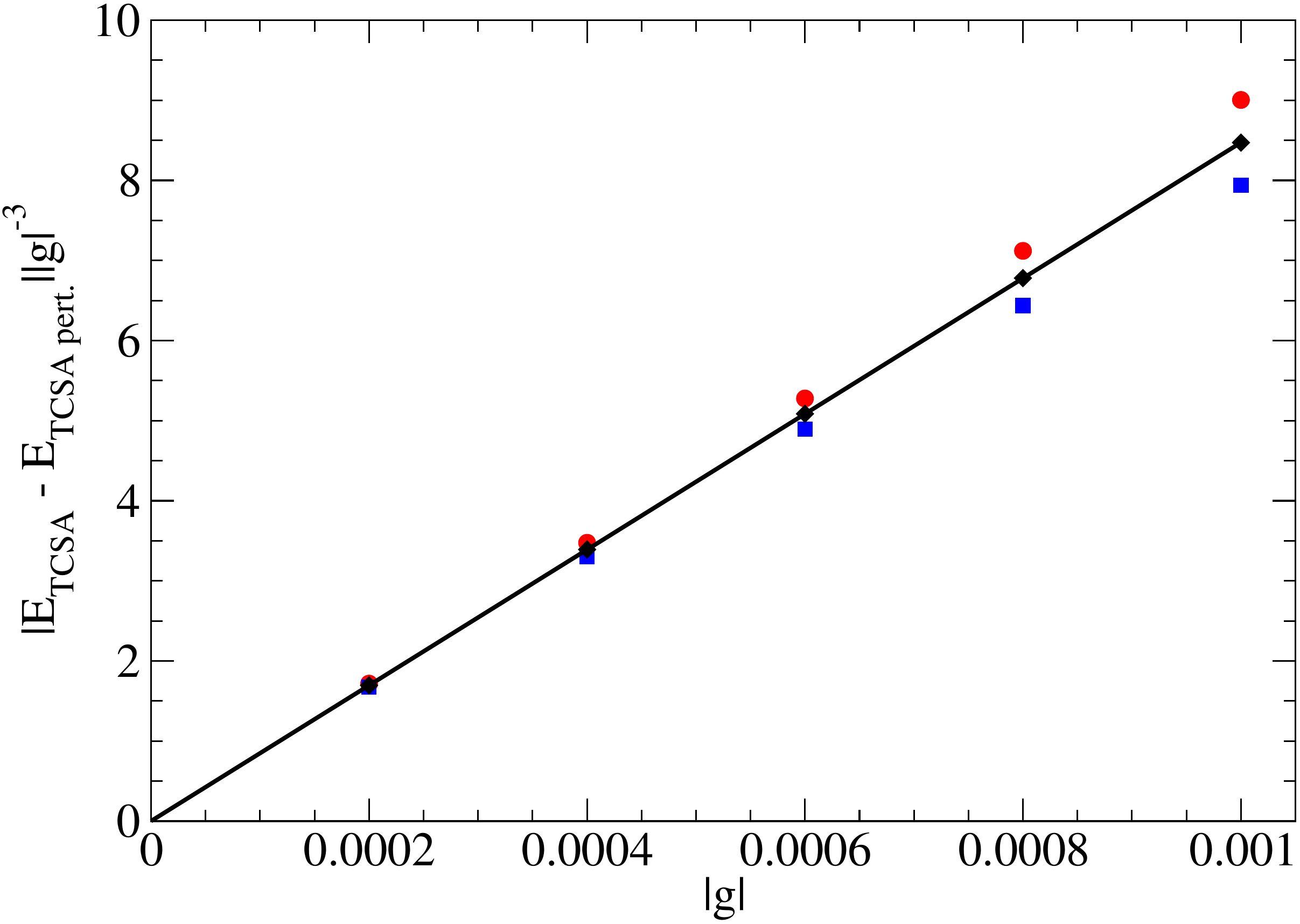}
\caption{Plots of the residual ground state energy (the TCSA data with the perturbative contributions, $I^{TCSA}_2$ and $I_3^{TCSA}$,
subtracted off) as a function of $g$ for  $N_{tr}=9$ (red: irrelevant $g$; 
blue: relevant $g$). The average of the two results 
(black curve) is fitted with  the function, $g^3(a+gbN_{tr}(N_{tr}+1))$ 
(this average removes lower order logs that might appear from resumming high order contributions).
The  fitting parameters are  $a=0.002\pm0.001$ and $b=94.12\pm0.02$.}
\label{marginal1}
\end{figure}

On the other hand, the universal corrections coming from $I_2$ and $I_3$  are
independent of the details of the regulator.  Such corrections are usually accessed  through analytic continuation in
the operator dimension $x$.
For sufficiently small $x$,
the resulting perturbative integrals become convergent.  For such a range of $x$, 
one then introduces a regulator, $\epsilon_p$, and expands in a Taylor series, finding
that $I_2$ and $I_3$ have the form \cite{Cardy,LudwigCardy}:
\begin{eqnarray}\label{I2I3x}
I_2(x,\epsilon_p) &=& c_2 \, \epsilon_p^{2(2-x)-2}+I_{2,univ.}(x)+{\cal O}(\epsilon_p^{6-2x});\cr\cr
I_3(x,\epsilon_p) &=& c_3 \, \epsilon_p^{3(2-x)-2} + I_{3,univ.}(x) + \epsilon_p^{2-x}I_{3,subleading}(x) +{\cal O}(\epsilon_p^{6-2x});\cr\cr
I_{3,subleading}(x) &=& -6\pi\frac{\epsilon_p^{2-x}}{2-x}  I_{2,univ.}(x).
\end{eqnarray}
Because everything is convergent, expressions can be developed for $I_{2,univ.}(x)$ and $I_{3,univ.}(x)$ as a function of $x$.
The universal terms' values close to $x=2$ are then the analytically continued parts of the expansion of $I_2$ and $I_3$ 
that are independent of the UV regulator (the non-universal terms, in contrast, in general either diverge or vanish close to $x=2$).
The relationship between $I_{3,subleading}(x)$ and $I_{2,univ}(x)$ arises from the OPE, $\phi\phi\sim b\phi$\cite{LudwigCardy}.

In  the case at hand, the universal contributions near $x=2$ are \cite{Cardy,LudwigCardy}
\begin{eqnarray}\label{I2I3}
I_{2,univ.}(x\sim 2) &=& -\frac{\pi}{4}(2-x);\cr\cr
I_{3,univ.}(x\sim 2) &=& -2\pi^2;\cr\cr
I_{3,subleading}(x\sim 2) &=& \frac{3\pi^2}{2}.
\end{eqnarray}
We also include the evaluation of $I_{3,subleading}(x)$ because
exactly at $x=2$ the prefactor of this term becomes independent
of $\epsilon_p$.  Its evaluation, with important consequences, will turn out to depend upon the choice of regulator.

The universal and subleading parts of $I_2$ and $I_3$ allow us to write the universal
part of $E_{0}$ solely as a function of the running coupling $\tilde g(l)$, compatible with scaling theory \cite{LudwigCardy}.  
$E_{0,univ}$ in terms of the bare dimensionless
coupling, $\tilde g  = g \epsilon_c^{2-x}$, is equal to
\begin{equation}
c(\tilde g) \equiv -\frac{6 R}{\pi}E_{0,univ}
=  c+(24\pi)\bigg(\frac{3}{8}(\epsilon_p^{-y}\tilde g)^2I_{2,\rm univ} + \frac{b}{6}(\epsilon_p^{-y}\tilde g)^3I_{3,univ})\bigg).
\end{equation}
Supposing $y\equiv (2-x)>0$, we can use the $\beta-$function to express $\tilde g$ in terms of $\tilde g(l)$:
\begin{equation}
\tilde g = \frac{\tilde g(l) \epsilon_p^{y}}{1-\frac{\tilde g(l)}{\tilde g^*}(1-\epsilon_p^y)},
\end{equation}
where $\tilde g^* = \frac{4\pi b}{3(2-x)} + {\cal O}(2-x)$ is the zero of the $\beta-$function.
We then have
\begin{equation}
c (\tilde g(l)) = c+24\pi\bigg(-\frac{3}{4}\frac{\pi(2-x)}{8}\tilde g^2(l) - \tilde g^3(l)(\frac{b\pi^2}{12} + {\cal O}(2-x)) + {\cal O}(\tilde g^4(l))\bigg).
\end{equation}
$c(\tilde g(l))$ is nothing more than Zamolodchikov's c-function \cite{ZamC}.  We see that $\partial_{\tilde g(l)}c(\tilde g(l))$ has the
same zero, $\tilde g^*$, as the $\beta-$function, as it should.  We also see that exactly at the marginal point, $x=2$, there
is a finite third order correction in $\tilde g(l)$ to the ground state energy.

An interesting question is what of this universal behavior can be extracted from the TCSA.  While it has been suggested in Ref. \cite{KlaMe2} that the dependency of the TCSA data upon the UV regulator (here, the truncation level, $N_{tr}$) obscures such terms, 
it has been shown in Ref. \cite{Watts2} that upon the subtraction of leading and sub-leading order divergences, universal IR behavior can be observed.  In answering these questions, the breaking of Lorentz invariance by the TCSA regulator will turn out to be key.

In order to determine the nature of the universal behaviour in the TCSA approach, 
we need to compute the integrals $I_2$ and $I_3$ with the TCSA regulator in place.  This however
can be straightforwardly done \cite{Watts2}.  The essential idea is that these integrals can be expanded in powers of the integration
variables $z_{1,2}$.  These expansions can then be compared with a Lehmann expansion of the corresponding correlation function
(see Appendix A.1).  By truncating
the Lehmann expansion to the same low energy states used in the TCSA, we are able to compute $I_{2,3}$ with the 
TCSA regulator.
The results precisely at the marginal point $x=2$ (see Appendices A.2 and A.3 for details) are surprisingly simple and can be computed exactly:
\begin{eqnarray}
I^{TCSA}_2 &=& \pi \, N_{tr}(N_{tr}+1);\nonumber\\
I^{TCSA}_3 &=& 3\pi^2 \, N_{tr}(N_{tr}+1). \label{pertur}
\end{eqnarray}
These expressions for $I_{2,3}$ contain both non-universal terms (as is evident from the dependency on $N_{tr}$) and potential universal
terms.  The question becomes how to identify which is which.  We do so by comparison with the evaluation of $I_2$ and $I_3$ in Refs. 
\cite{Cardy,LudwigCardy} already given in Eqn.(\ref{I2I3}).  At $x=2$, $I_2(x=0)$ 
is purely non-universal, i.e. $I_{2,univ.}(x=2)=0$.  This corresponds to our finding
that $I_2$ is proportional to $N_{tr}(N_{tr}+1)$ and indicates that the regulator, $\epsilon_p$, used in Ref. \cite{Cardy,LudwigCardy} can
be identified with the TCSA regulator via $\epsilon_p^{-2} \propto N_{tr}(N_{tr}+1)$.  However  $I^{TCSA}_3$ is also 
proportional to $N_{tr}(N_{tr}+1)$, which seems to imply that the third order contribution to $E_{gs}$, as with second order, is purely
non-universal.  We see then already at $x=2$ that the TCSA regulator leads to a different universal structure to $E_{0,univ}$ 
than the Lorentz invariant regulator employed in \cite{Cardy,LudwigCardy}.

We verify the accuracy of this perturbative computation by comparing it with the TCSA numerics.  In Fig.\ref{marginal}  
we plot $E_0$ at small $g$ evaluated numerically against the perturbative results.  We obtain excellent agreement.
To analyze more closely the possible presence of a universal term in the numerics, we plot in Fig. \ref{marginal1} the residual ground state energy
arrived at by subtracting from the numerical data the perturbative contributions (solely non-universal) 
corresponding to $I^{TCSA}_{2}+I^{TCSA}_3$.  To determine
whether the numerics indicate any universal contribution, we plot the residual as a function of g and fit the results to a function of the
form $g^3(a+gbN_{tr}(N_{tr}+1))$.  These fits put an approximate bound (the value of $a$) 
on the third order universal term consistent with the numerics.  
We find it to be 
considerably smaller than that found in Ref. \cite{Cardy,LudwigCardy}, consistent with our previous statement that in the regulation scheme
used by the TCSA, there is no universal term at third order in the coupling.

To get at the origin of the discrepancy between the TCSA evaluation of $E_{0,univ}$ and that of Refs. \cite{Cardy,LudwigCardy},
we evaluate the integrals, $I_2$ and $I_3$ in the TCSA regulation scheme away from the marginal point.  These
integrals have the structure (compare with Eqn. \ref{I2I3x})
\begin{eqnarray}
I_2^{TCSA}(x,N_{tr}) &=& I^{TCSA}_{2,div.}(x)(N_{tr}(N_{tr}+x-1))^{x-1} + I^{TCSA}_{2,univ.}(x) + I^{TCSA}_{2,subleading}N_{tr}^{2x-6};\cr\cr
I_3^{TCSA}(x,N_{tr}) &=& I^{TCSA}_{3,div.}(x)(N_{tr}(N_{tr}+x-1))^{x-1} + I^{TCSA}_{3,univ.}(x) + I^{TCSA}_{3,subleading}N_{tr}^{2x-4}.
\end{eqnarray}
We can evaluate the divergent and universal parts exactly (see Appendices A.4 and A.5)
\begin{eqnarray}\label{evalI3}
I^{TCSA}_{2,div.}(x) &=& \frac{2\pi}{\Gamma^2(x)(2x-2)};\cr\cr
I^{TCSA}_{2,univ.}(x) &=& \frac{\pi\Gamma^2(\frac{x}{2})\Gamma(1-x)}{\Gamma^2(1-\frac{x}{2})\Gamma(x)};\cr\cr
I^{TCSA}_{3,div.}(x) &=& \frac{12\pi^2}{\Gamma(\frac{x}{2})^6}\int^1_0dj\int^j_0d\bar j \int^{1-j}_0 dl\int^{1-j}_0 dk \frac{(j\bar j k l (l+j-\bar j)(k+j-\bar j))^{\frac{x}{2}-1}}{(j+l)(k+j)}
;\cr\cr
I^{TCSA}_{3,univ.}(x) &=& -\frac{2-x}{4-3x}\frac{1}{2}(L_{11}L_{12}+L_{21}L_{22});\cr\cr
L_{11} &=& \frac{\sqrt{\pi}}{2^{x}\sin (\frac{\pi x}{2})}\frac{\Gamma (1-\frac{x}{2})\Gamma(1-\frac{3x}{4})\Gamma(\frac{3}{2}-\frac{x}{2})\Gamma(\frac{x}{4})\Gamma (\frac{x}{2})}{\Gamma(\frac{x}{2})\Gamma(2-x)\Gamma^2(1-\frac{x}{4})};\cr\cr
L_{12} &=& -\frac{\pi (x-2) B(\frac{x}{2},\frac{x}{2})}{2\sin (\frac{\pi x}{2})}~ {_3F}_2(2-\frac{x}{2},\frac{x}{2},\frac{x}{2},2,x,1);\cr\cr
L_{21} &=& L_{11};\cr\cr
L_{22} &=& B(\frac{x}{2},\frac{x}{2}-1)\Gamma(\frac{x}{2})\Gamma(1-\frac{x}{2}) ~{_3F}_2(\frac{x}{2},\frac{x}{2}-1,1-\frac{x}{2},x-1,1,1),
\end{eqnarray}
where $B(x,y)$ is the $\beta-$function and $_3F_2$ is a generalized
hypergeometric function.  The subleading terms we are only able to evaluate numerically.  Close to $x=2$ we find
\begin{eqnarray}
I^{TCSA}_{2,subleading}(x=2) &=& 0;\cr\cr
I^{TCSA}_{3,subleading}(x=2) &=& 2\pi^2.
\end{eqnarray}
Plots of their values near $x=2$ can be found in Appendices A.4 and A.5.

Now how does this compare to the results of \cite{Cardy,LudwigCardy}? Remarkably
the universal (constant) parts of $I_2$ and $I_3$
are the same (compare Eqn. (\ref{I2I3x})):
\begin{eqnarray}
I^{TCSA}_{2,univ.}(x) &=& I_{2,univ.}(x);\cr\cr
I^{TCSA}_{3,univ.}(x) &=& I_{3,univ.}(x).
\end{eqnarray}
In particular
\begin{eqnarray}
I^{TCSA}_{2,univ.}(x\sim 2) &=& -\frac{\pi}{4}(2-x);\cr\cr
I^{TCSA}_{3,univ.}(x\sim 2) &=& -2\pi^2.
\end{eqnarray}
We however see discrepancies in the subleading term,
$I_{3,subleading}$:
\begin{equation}
I_{3,subleading}(x\sim 2) =\frac{3\pi^2}{2}; ~~~~ I^{TCSA}_{3,subleading}(x\sim 2) = 2\pi^2.
\end{equation}
This difference is a consequence of the TCSA's different (non-Lorentz
invariant) regulator.  It is possible to understand this difference as $I_{3,subleading}$ is finite at $x=2$ due to a cancellation in
a pole term due to an OPE and a zero in $I_{2,univ.}$.  This delicate cancellation leaves $I_{3,subleading}$ sensitive
to choice of regulator in a way that $I_{3,univ.}$ is not.

What are then the implications of $I_{3,subleading}(x)$ depending upon the regulator?  Firstly we obtain an
altered Zamolodchikov c-function:
\begin{equation}
c_{TCSA} (\tilde g(l)) = c+24\pi\bigg(-\frac{3}{4}\frac{\pi(2-x)}{8}\tilde g^2(l) - \tilde g^3(l)(\frac{b\pi^2}{12}(1-\epsilon_{TCSA}^{2-x}))\bigg)
\end{equation}
where $\epsilon_{TCSA} = N_{tr}^{-1}$.  Interestingly the TCSA c-function agrees with that derived previously in Ref. \cite{LudwigCardy}
for $x<2$ in the large $N_{tr}$ (small $\epsilon_p$) limit.  However it is not solely a function of $\tilde g(l)$ as would be
suggested by scaling theory.  

As we have already noted, at $x=2$ the $\tilde g^3 (l)$ term vanishes in $c_{TCSA}(\tilde g)$.  
This suggests that corrections that contribute to the universal part of the ground state energy are sensitive to the use of
a non-Lorentz invariant regulator.  It is interesting to note that
these Lorentz invariant corrections are seen in models which are the {\it lattice} equivalent of these field theories and
so do not have Lorentz invariance.
Ref. \cite{Kluemper}, using a quantum transfer matrix approach,
has analyzed the XXX Heisenberg model, a lattice equivalent of
$SU(2)_1$ with a marginally irrelevant current-current interaction.
In this work it is shown that the $g^3$ term as computed in
Refs. \cite{LudwigCardy,Lukyanov} using a Lorentz-invariant
regulator is seen in the low temperature specific heat of the spin
chain.  At least for the XXX spin chain, terms in its field theoretic equivalent Hamiltonian that break Lorentz invariance
are irrelevant rather than marginal in nature \cite{Lukyanov}.

An interesting possibility \cite{CardyPC} is that the universal structure of the ground state energy as determined
in a Lorentz invariant field theory remains in the presence of a non-Lorentz invariant regulator provided one allows
the ``speed of light'' in the system to be renormalized by the interactions (a possibility if Lorentz invariance is broken), that is
to say that all of the consequences of breaking Lorentz invariance reside in a renormalization of $c$.  Because the ground
state energy depends upon $1/c$ (throughout this paper we have set the bare value of $c$ to 1), we cannot distinguish
between a breaking of universality in the ground state energy and a renormalization of $c$. 
\subsection{NRG and Marginal Perturbations}

\begin{figure}[t]
\centering
\includegraphics[scale=0.2]{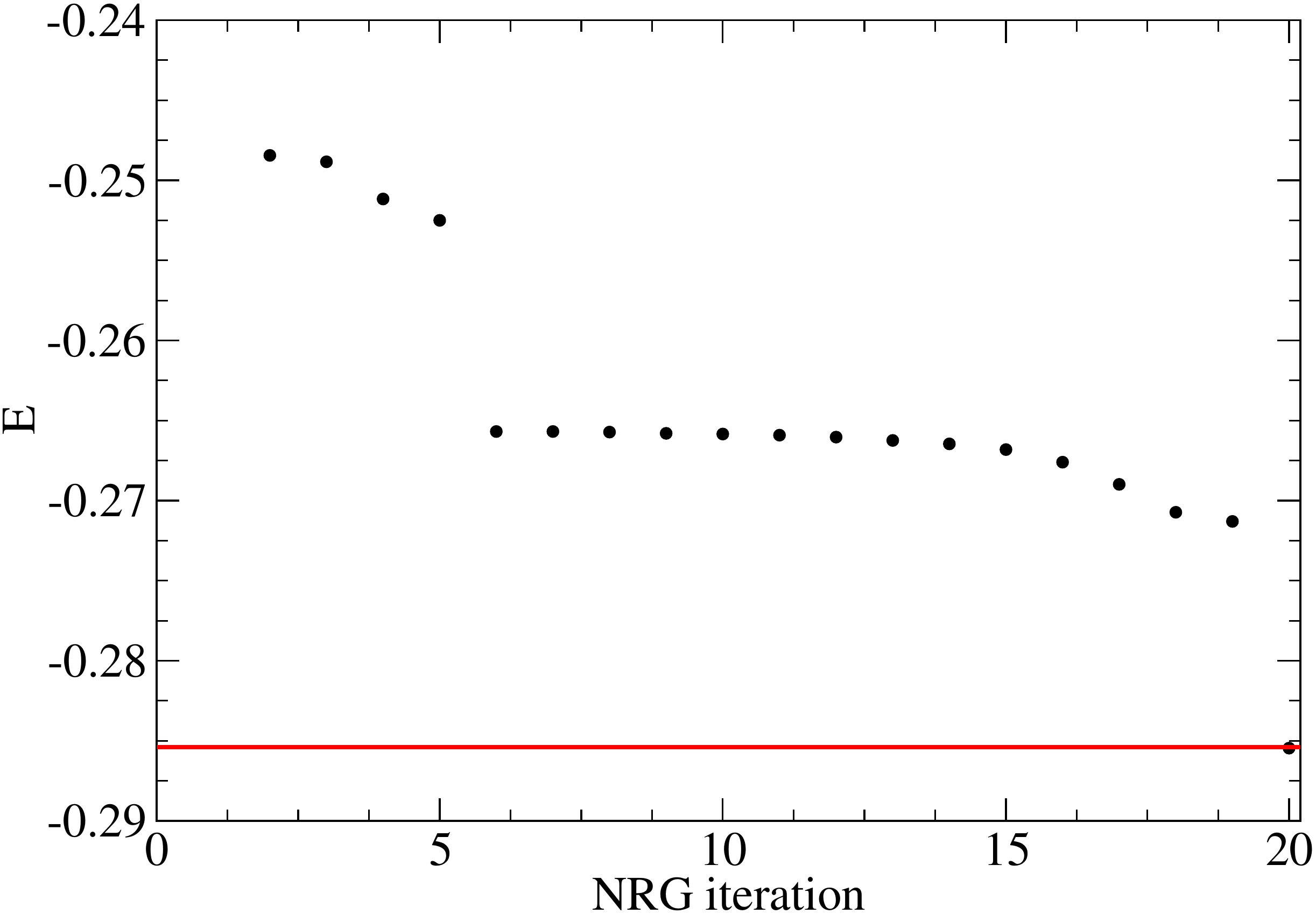}\includegraphics[scale=0.2]{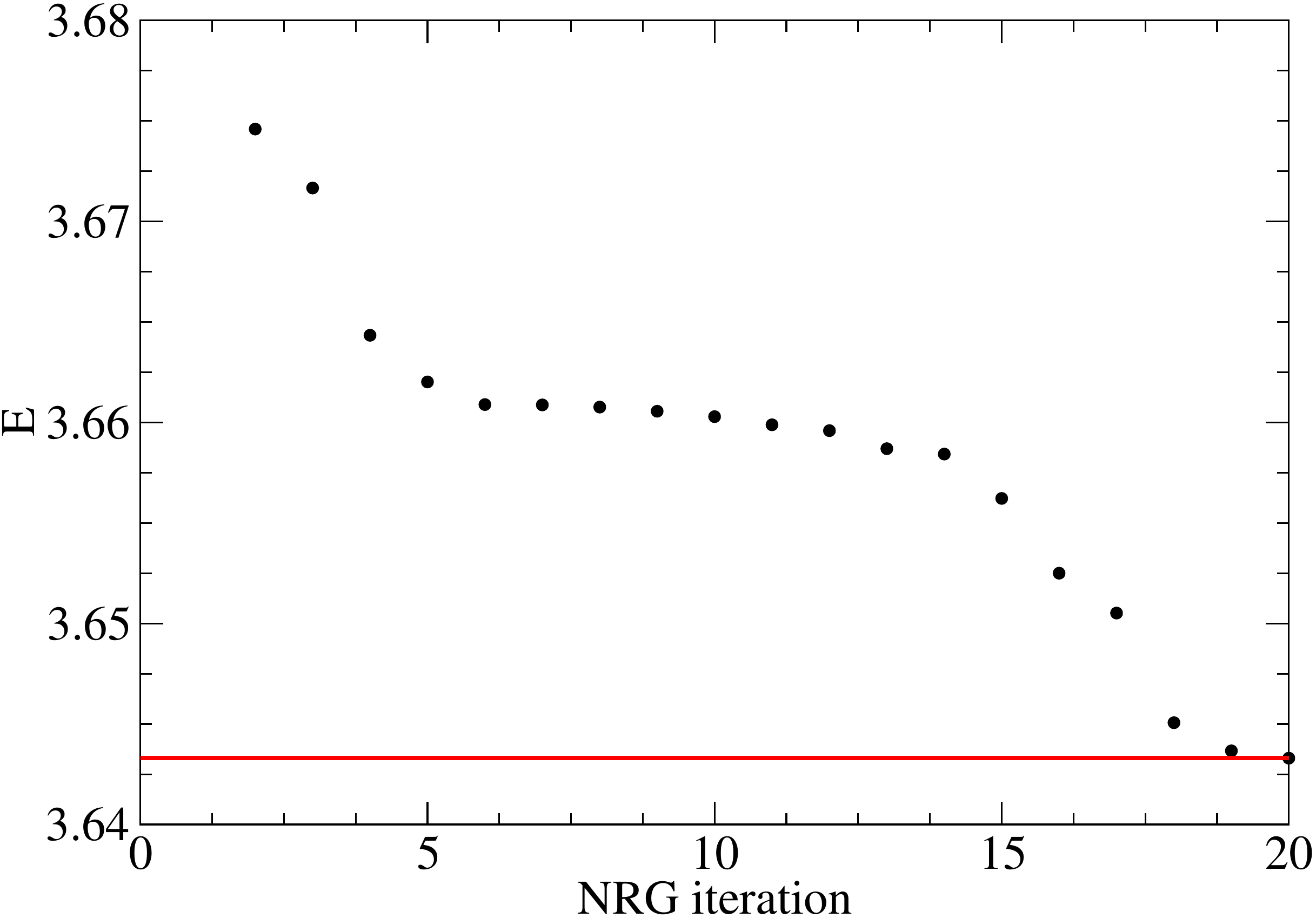}\includegraphics[scale=0.2]{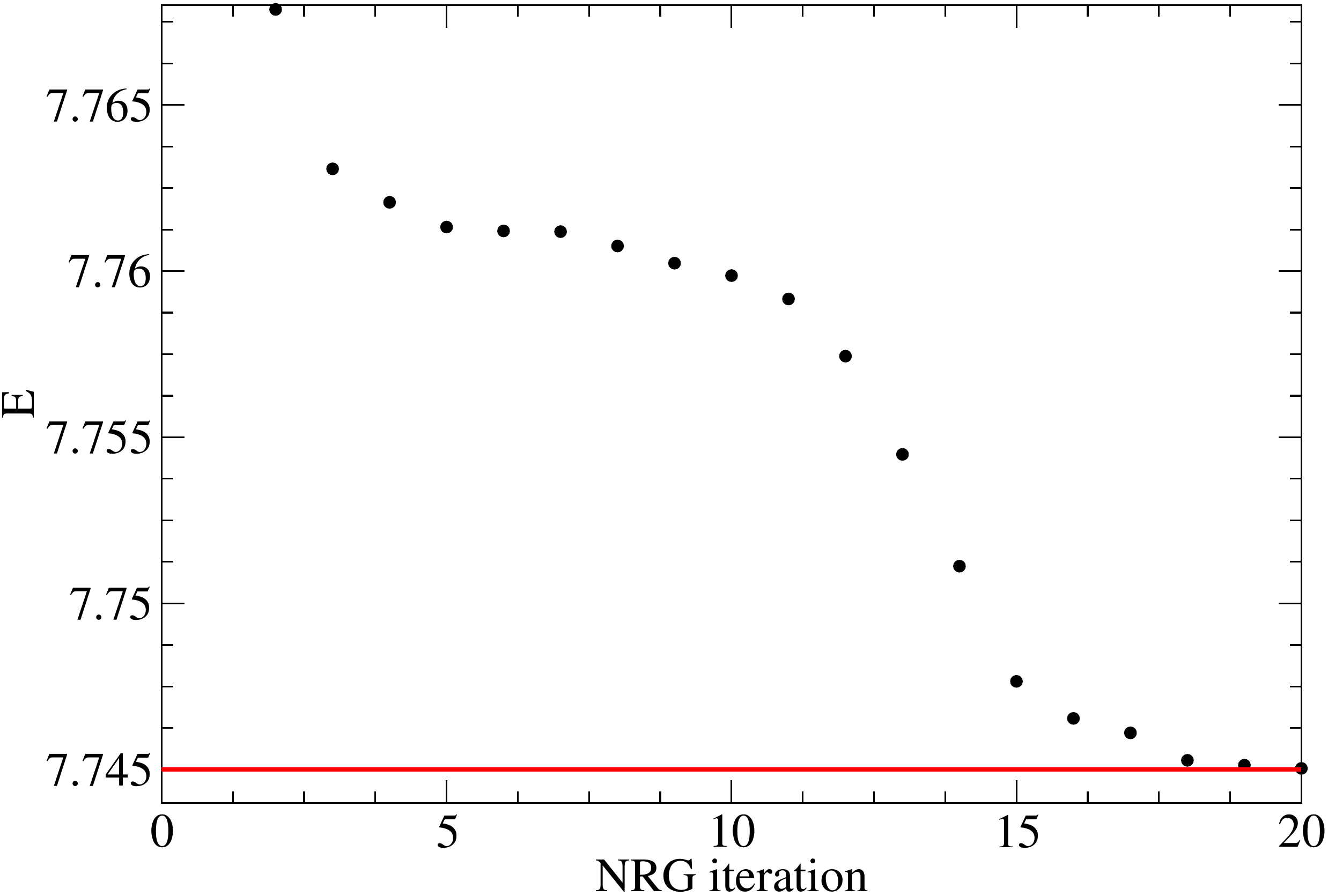}
\caption{ 
Plots of the ground state (left), the first excited state (center), and the fifth excited state (right) energies as a function of RG step.
The energies of the last RG step correspond to taking into account all states up to a truncation of $N_{tr}=11$.  Here
$g=0.008$ is marginally irrelevant. The NRG
base matrix size and step size ($N$ and $\Delta$ in the notation of \cite{NRG1}) are $N=4767$ and $\Delta=500$.
The total number of states in the Verma module of the Identity is 14767.
The solid line is the TCSA result done with an exact diagonalization at level $N_{tr}=11$. 
}
\label{nrg+marginal}
\end{figure}

In this subsection we apply the numerical renormalization group to the study of the marginal current-current perturbation
of $SU(2)_1$.  The NRG is a technique that allows the TCSA to include states at much higher conformal levels than would be possible 
with a straight exact diagonalization.  It does so by taking a cue from Kenneth Wilson's NRG \cite{wilson}: it takes into account the states
that have a weaker influence on the low energy eigenstates of the full theory, in this case high energy conformal states, only in numerically
manageable chunks.  It works in the case of a relevant perturbation because such perturbations guarantee that the high energy conformal
Hilbert space only affects  weakly  the low energy sector of the theory.  Thus it is not clear, {\it a priori}, whether the NRG will work 
in the case of a marginal perturbation where the high and low energy sectors of the theory are more tightly coupled.  
We will, however, see that the NRG does work, reproducing with high accuracy the results of the TCSA run with a straight exact diagonalization.

Fig. \ref{nrg+marginal} shows  the RG evolution of the energies of  the ground state, the first excited state, 
and the fifth excited state.   The NRG results
converge towards  the exact diagonalization results  with excellent accuracy.  At least for the low lying energies in marginally perturbed conformal
field theories, the NRG seems to be able to reproduce the expected energies, despite their dependencies upon the TCSA UV cutoff, $N_{\rm tr}$.

\begin{figure}
\centering
\includegraphics[scale=0.4]{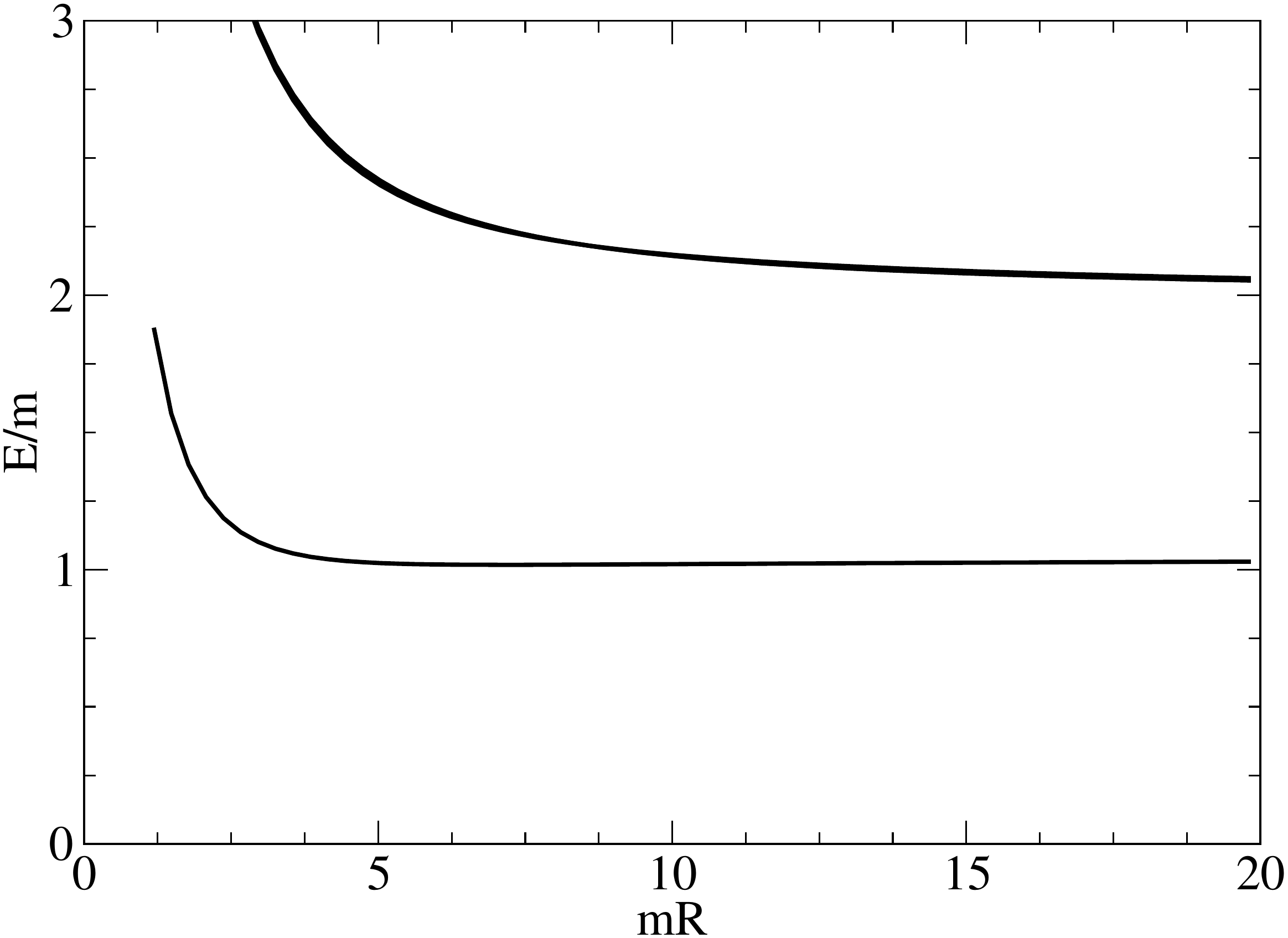}
\caption{Plot of the energies of the four lowest excited states as a function of $R$ in the sector $S_z=0$. 
The ground state energy has been subtracted. 
The data are found  for a truncation level $N_{tr}=6$. There is  a single  low lying state with mass $M=1.001$ 
and  spin $S_z=0$,  which is a member  of the fundamental  triplet of particles. The higher energy level   is  three fold degenerate 
and corresponds to two-particle states.}  
\label{su2spectrum}
\end{figure}

\section{ $SU(2)_2$ perturbed by the spin-1 field}
\label{lev2}
In this section we apply the TCSA to a deformation of the $SU(2)_2$ WZW model \cite{dif}. 
The $SU(2)_2$ WZW model is a $c=\frac{3}{2}$ conformal 
field theory with three primary fields, the spin-0 identity field, a spin $\frac{1}{2}$ field, $\phi_{1/2,\pm 1/2}$, 
and a spin $1$ field, $\phi_{1,\{\pm 1,0\}}$, with conformal 
weights $\Delta=0$, $\Delta_{1/2}=\frac{3}{16}$, and $\Delta_1=\frac{1}{2}$ respectively.

$SU(2)_2$ WZW can be thought of as a theory of three non-interacting massless Majorana fermions (the spin-1 fields).
Perturbing $SU(2)_2$ with the spin-1 field,
\begin{equation}
 H=H_{SU(2)_2}+g\int dx\left( \phi_{1,1} \, \bar\phi_{1,-1}-\phi_{1,0} \, \bar\phi_{1,0}+\phi_{1,-1} \, \bar\phi_{1,1}\right),
\label{Ham3ferm}
\end{equation}
makes the fermions massive.  Finding such a spectrum however is a strong check on the TCSA as the massive Majorana
fermions are not simply expressible in the conformal current algebra basis.

\begin{figure}
 \centering
\includegraphics[scale=0.3]{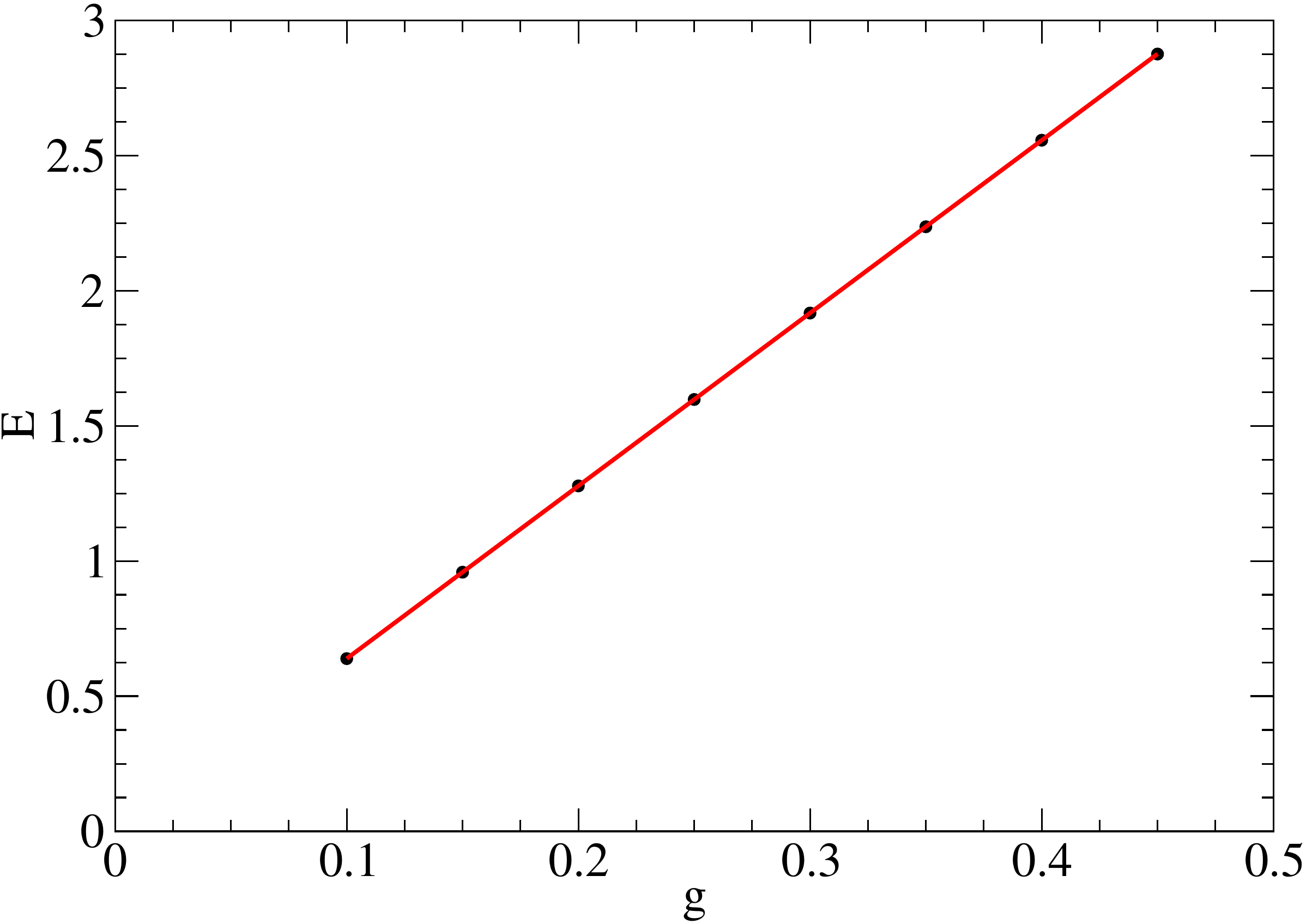}\includegraphics[scale=0.3]{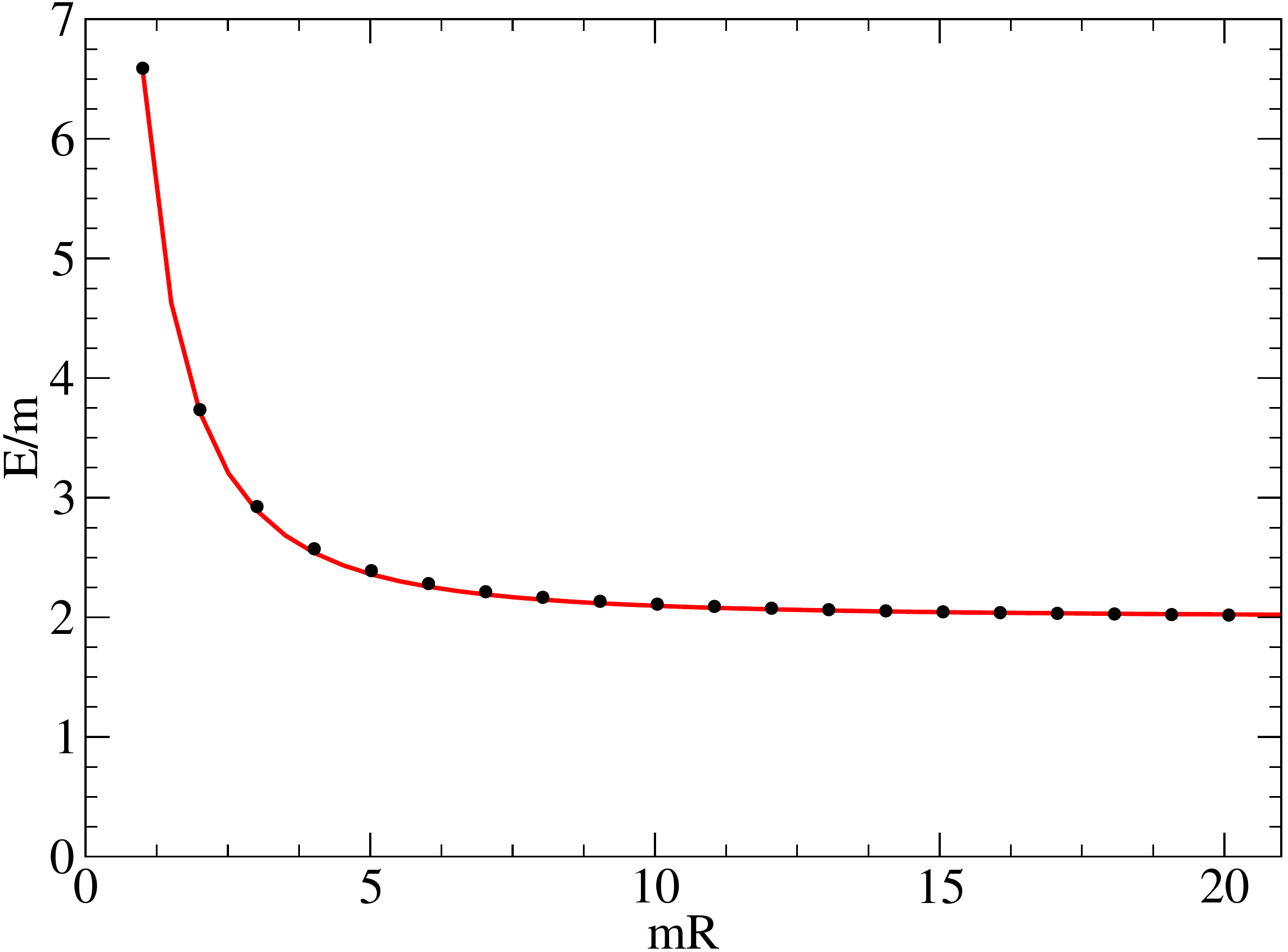}
\caption{a)  Linear scaling of the mass of the triplet with the coupling $g$.  
b) A comparison of the energy of a two-particle state with $(n_1,n_2)=(-1,1)$
between TCSA numerics (black dots) and analytics (solid red line).}
\label{su2_2}
\end{figure}

Fig. \ref{su2spectrum} shows  the low lying excited states from the TCSA.  The coupling $g$ is fixed to a value
where the mass of the first excited state is   1.  The TCSA captures this state (the $S_z=0$ state
of the triplet of massive Majoranas) as well as a set of three two particle states with an energy approximately equal to 2.  
The Majorana fermions are non-interacting, hence 
their mass must  scale linearly with $g$.  This behavior is shown in  Fig. \ref{su2_2}a.   We also
consider the finite size corrections to one of the two-particle states.  Even though non-interacting, the quantization
condition for the two-particle state, Eqn. \ref{quant}, is non-trivial as one needs  to quantize with
a doublet of distinct integers $(n_1,n_2)$, $n_1\neq n_2$.   Fig. \ref{su2_2}b  shows that  
the TCSA energy of the two-particle state matches the prediction derived from  Eqn. \ref{quant}.

\section{Conclusions and Discussions}
In this paper we have applied the TCSA to perturbations of $SU(2)_k$.
The general  methodology is illustrated with various  examples.  We began by studying the 
$SU(2)_1$ + $Tr (g)$ perturbation, equivalent to the sine-Gordon model at a particular value of its
coupling.  We showed the TCSA accurately captured both the ground state energy and excited state spectrum
including finite size corrections.  We next turned to $SU(2)_1$ + $\bar{J}_L\cdot\bar{J}_R$, a WZW model
perturbed by a marginal interaction.   We demonstrated that one can
accurately identify analytically the leading UV divergences in perturbation theory that characterize the ground state energy as computed 
numerically by the TCSA.  By subtracting these UV divergences
(i.e. the non-universal contributions to the ground state energy)
we isolated the universal contribution to the ground state energy due to the marginal perturbation.  We find that it
differs from that predicted with calculations using Lorentz invariant regulators.  Interestingly, away from the marginal
point, the universal structure of the ground state energy is restored.
Finally we considered $SU(2)_2$ perturbed by $Tr(g)^2$.  This model is equivalent to three massive non-interacting
Majorana fermions.  The TCSA, is able to reproduce the  expected spectrum.  This is a non-trivial check of our methodology 
as the Majorana fermions do not have a simple representation in the current algebra basis employed by the approach.

We have developed this capability to study perturbed WZW models in order to tackle a number of problems.  In particular, we have two in mind.
In the first, we plan to examine the dimerized-frustrated   $J_1-J_2- \delta$ Heisenberg model whose Hamiltonian is 
\begin{equation}
H=\sum_n J_1\, (1+\delta(-1)^n) \, \bar{S}_n\cdot\bar{S}_{n+1} + J_2 \, \bar{S}_n\bar{S}_{n+2}.
\end{equation}
Field theory analyses \cite{Af1}
argue that the low energy sector of this theory is equivalent to
\begin{equation}
H = H_{SU(2)_1} + g\int dx \, \bar{J}_R\cdot\bar{J}_L + h\int dx \, (\phi_{1/2,1/2} \, \bar\phi_{1/2,-1/2}-\phi_{1/2,-1/2} \, \bar\phi_{1/2,1/2}),
\end{equation}
where $g \propto J_2-J_{2c}$ and $h \propto \delta$.  If the marginal perturbation $g$ is  absent, the model's spin gap, $\Delta_S$, would
scale simply with $h$: $\Delta_S \propto h^{2/3}$.  In the presence of $g$, however, this scaling is altered to become
$\Delta_S \propto  h^{2/3}/|\log (h)|^{1/2}$.  However this altered scaling has been difficult to see in DMRG
studies of the dimerized-frustrated   Heisenberg model \cite{branes,kumar}.  It would be extremely interesting to analyze 
this scaling behavior using the TCSA.  
 
Our study here of $SU(2)_1$ + $\bar{J}_L\cdot\bar{J}_R$ has thus lain the groundwork
for this future study.

The second problem that we intend to tackle is the study of possible integrable perturbations of $SU(2)_k$ for $k>1$.  There are 
indications,  coming from  Zamolodchikov-type 
counting arguments \cite{Zamcounting},    that those  perturbations do exists.  We intend to study these perturbations
with the TCSA, extracting both the spectrum of the model as well as evidence for or against their integrability.  The
example analyzed in this paper concerning $SU(2)_2$ shows that this goal is within reach.

\acknowledgments
We thank J. Cardy, F. H. L. Essler, G. Mussardo and G. Takacs for useful discussions. L.L. acknowledges a grant awarded by Banco de Santander and financial support from European Regional Development Fund.
R.M.K. acknowledges support by the US DOE under contract DE-AC02-98CH10886 and NSF
under grant no. PHY 1208521. G.S. acknowledges support from  the grants 
FIS2009-11654, QUITEMAD and the Severo-Ochoa Program.
G.P.B. acknowledges support from the Netherlands Organisation for Scientific Research (NWO).

\appendix

\section{Evaluation of the Perturbative Integrals from the Current-Current Perturbation of $SU(2)_1$}

\subsection{TCSA regularization and the Lehmann representation}

We implement the $UV$ regularization explicit in the TCSA with a truncation of the Lehmann representation of 
the $n$-point functions \cite{Watts1,Watts2}. As an explicit example, we compute the two point function using this truncation.  Introducing
projectors $\mathcal P_{N_{tr}}$ and $\bar\mathcal { P}_{N_{tr}}$ that project out 
all chiral states of level higher than $N_{tr}$, we can write the two point function appearing in the evaluation
of the second order perturbative contribution to the ground state energy as
\begin{equation}
\langle{\cal T} \, \phi(x,t)\phi (0,0)\rangle = \langle \mathcal {T} \, \mathcal{P}_{N_{tr}}\bar\mathcal{P}_{N_{tr}}\phi(x,t)\mathcal{P}_{N_{tr}}
\bar\mathcal{ P}_{N_{tr}}\phi(0,0)\mathcal{P}_{N_{tr}}\bar\mathcal{ P}_{N_{tr}}\rangle,
 	\label{p}
\end{equation}
where $\mathcal T$ is the usual time ordering operator. These projectors act to truncate the Lehmann mode expansion
arising from the insertion of the resolution of the identity in between the fields of (\ref{p}):
\begin{equation}
\sum_{0\leq m,\bar{m}\leq N_{tr}}\langle 0|\phi(0,0)|m\rangle\otimes|\bar{m}\rangle\langle m|\otimes\langle\bar{m}|\phi(x,t)|0\rangle.
\end{equation}
Here the truncated sum is over all states whose conformal level is less than $N_{tr}$ and we just considered the contribution with $t<0$.
Using the time and space translation operator, $e^{-Ht-iPx}$, and defining 
$z=e^{\frac{2\pi}{R}(ix+t)}$ and $\bar z=e^{\frac{2\pi}{R}(-ix+t)}$, we can rewrite the above as
\begin{eqnarray}
\sum_{0\leq m,\bar{m}\leq N_{tr}}\langle 0|\phi(0,0)(|m\rangle\otimes|\bar{m}\rangle\langle m|\otimes\langle\bar{m}|)\phi(0,0)|0\rangle z^m\bar{z}^{\bar m},
\label{leh2}
\end{eqnarray}
so reducing the correlation function to a sum over powers in $z$ and $\bar z$.  We see in this
sum that no power greater than $N_{tr}$ of $z$ or $\bar z$ appears.

The case of the third order correction to the energy involving a three-point function follows the same strategy. 
We insert the projectors 
\begin{equation}
 \langle \mathcal {T} \, \mathcal{P}_{N_{tr}}\bar\mathcal{P}_{N_{tr}}\phi(x,t)\mathcal{P}_{N_{tr}}
\bar\mathcal{ P}_{N_{tr}}\phi(0,0)\mathcal{P}_{N_{tr}}\bar\mathcal{ P}_{N_{tr}}\phi(x',t')\mathcal{P}_{N_{tr}}\bar\mathcal{ P}_{N_{tr}}\rangle,
 	\label{p3}
\end{equation}
arriving, in the case where $t'<0<t$, at
\begin{equation}
\sum_{0\leq m,\bar{m},n,\bar{n}\leq N_{tr}}\langle 0|\phi(x,t)|m\rangle\otimes|\bar{m}\rangle\langle m|\otimes\langle\bar{m}|\phi(0,0)|n\rangle\otimes|\bar{n}\rangle\langle n|\otimes\langle\bar{n}|\phi(x',t')|0\rangle.
\end{equation}
If we again use the space time translation operator and use the notation $z=e^{\frac{2\pi}{R}(ix+t)}$, $\bar z=e^{\frac{2\pi}{R}(-ix+t)}$,
$w=e^{\frac{2\pi}{R}(ix'+t')}$, and $\bar w=e^{\frac{2\pi}{R}(-ix'+t')}$, we obtain
\begin{equation}
\sum_{0\leq m,\bar{m},n,\bar{n}\leq N_{tr}}\langle 0|\phi(0,0)|m\rangle\otimes|\bar{m}\rangle\langle m|\otimes\langle\bar{m}|\phi(0,0)|n\rangle\otimes|\bar{n}\rangle\langle n|\otimes\langle\bar{n}|\phi(0,0)|0\rangle z^{-m}\bar{z}^{-\bar m} w^{n} \bar{w}^{\bar{n}},
\label{leh3}
\end{equation}
that is a truncated polynomial at order $N_{tr}$ in $z,\bar z,w,\bar w$.

\subsection{Evaluation of $I_2$ at $x=2$}

We want to compute the integral, $I_2^{TCSA}$, by regularizing as we did with (\ref{p}).  Transforming the integral in $I_2$ back to the cylinder
we obtain,
\begin{equation}
I_2^{TCSA} = 2\cdot \left(\frac{2\pi}{R}\right)^2 \int_0^R dx\int_{-\infty}^0  dt \, \frac{z}{(z-1)^2}\frac{\bar{z}}{(\bar{z}-1)^2},
\end{equation}
where the factor of two counts the two contributions coming from time ordering.
If we then expand the integrand as a power series in $z,\bar{z}$ and compare with Eqn. \ref{leh2}, we see we should truncate the sum as 
follows 
\begin{equation}\label{I2}
I_2 = 2\cdot \left(\frac{2\pi}{R}\right)^2\int_0^R dx\int_{-\infty}^0  dt \, \sum_{n=0}^{N_{tr}}nz^n \, \sum_{n=0}^{N_{tr}}m\bar{ z} ^m ,
\end{equation}
whereupon the integral and sum are easily computed:
\begin{equation}
I_2 = \pi N_{tr}(N_{tr}+1).
\end{equation}
Comparing with the evaluation in \cite{Cardy,LudwigCardy}, we see that we obtain a relationship between the TCSA regulator and the
short distance cutoff, $\epsilon_p$, used in \cite{Cardy,LudwigCardy}:
\begin{equation}
\epsilon_p=\frac{1}{\sqrt{N_{tr}(N_{tr}+1)}}.
\label{epsilon}
\end{equation}

\subsection{Evaluation of $I_3$ at $x=2$}

We now turn to computing $I_3^{TCSA}$:
\begin{equation}
I_3^{TCSA}=\int dzd\bar z dw d\bar w \, \frac{1}{(z-w)(\bar z-\bar w)(z-1)(w-1)(\bar z -1)(\bar w-1)}.
\end{equation}
Moving to the cylinder, the time ordering gives $6$ contributions, $t<t'<0$, $t>t'>0$, $t<0<t'$ 
and those with $t\leftrightarrow t'$, all being equal. We show, as an example, the integral for $t>0>t'$:
\begin{equation}
\left(\frac{2\pi}{R}\right)^4\int_0^R dx \int_0^R dx' \int_0^\infty dt \int_{-\infty}^0 dt' 
\frac{z\, \bar z \, w \, \bar w}{(z-w)(\bar z-\bar w)(z-1)(w-1)(\bar z -1)(\bar w-1)},
\label{3rd}
\end{equation}
where $z=e^{\frac{2\pi}{R}(ix+t)}$, $\bar z=e^{\frac{2\pi}{R}(-ix+t)}$, $w=e^{\frac{2\pi}{R}(ix'+t')}$, and $\bar w=e^{\frac{2\pi}{R}(-ix'+t')}$.
We compute the contribution this makes by expanding the above in a power series of $z,\bar z,w,\bar w$ 
and truncating it at the $N_{tr}$-th power, matching the truncation in the Lehmann expansion in Eqn. \ref{leh3}. 
The integrand in (\ref{3rd}) then becomes
\begin{eqnarray}
\frac{|z|^2 |w|^2}{|z-w|^2|z-1|^2|w-1|^2}&=&\frac{1}{|1-w/z|^2}\frac{1}{|1-1/z|^2}\frac{1}{|1-w|^2}\frac{|w|^2}{|z|^2}\\
&=&\sum_{j,k,l\bar j,\bar k,\bar l\in C}\left(\frac{z}{w}\right)^{j+1}\left(\frac{\bar z}{\bar w}\right)^{\bar j+1} z^{-k} \, \bar z ^{-k} \, w^l \, \bar w^l\nonumber\\
&=&\sum_{j,k,l\bar j,\bar k,\bar l\in C} w^{j+l+1} \bar w ^{\bar j +\bar l+1} z^{-k-j-1} \, \bar z ^{-\bar k-\bar j -1},
\end{eqnarray}
where $C$ is defined by
  \begin{eqnarray}
    & &1\leq j+l+1\leq N_{tr};\nonumber \\
    & &1\leq j+k+1\leq N_{tr};\nonumber \\
    & &1\leq \bar j+\bar l+1\leq N_{tr}; \nonumber \\
    & &1\leq \bar j+\bar k+1\leq N_{tr}.
\end{eqnarray}
and all the indexes are positive or zero. The integrals can now be done and the sum that so arises evaluated to be 
\begin{equation}
\pi^2\sum_{C}\frac{1}{(j+l+1)(j+k+1)}  = \pi^2 \, \frac{N_{tr}(N_{tr}+1)}{2} .
\end{equation}
Once combined with the other five equal contributions, we obtain the final result:
\begin{equation}
 I_3^{TCSA}=3\pi^2N_{tr}(N_{tr}+1).
\end{equation}

\subsection{Evaluation of $I^{TCSA}_2(x)$}

\begin{figure}
\centering
\includegraphics[scale=0.4]{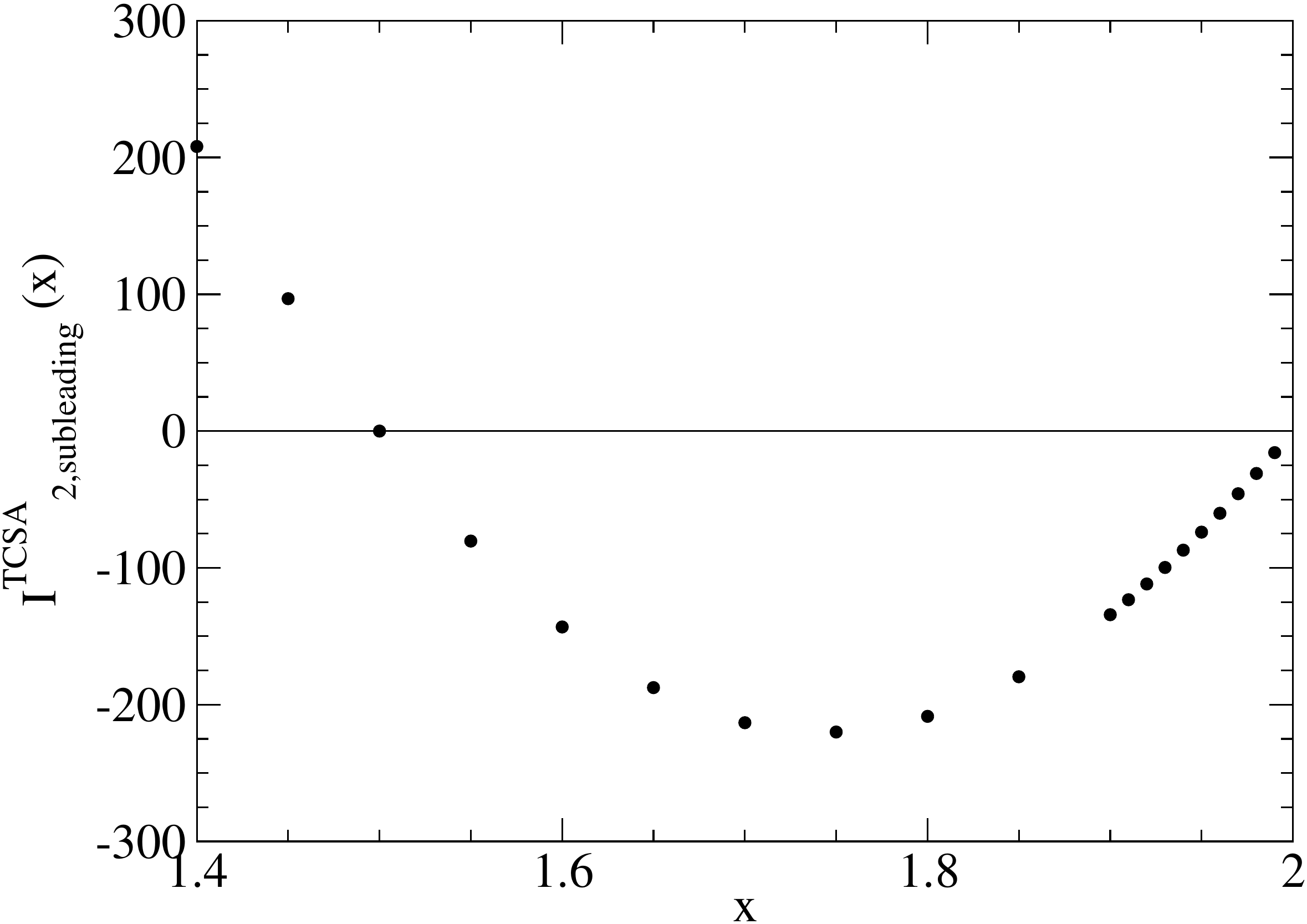}
\caption{Plot of  $I^{TCSA}_{2,subleading}(x)$ as a function of x.}
\label{I2subleading}
\end{figure}

For general $x$, $I_2$ once transformed back onto the cylinder reads
\begin{equation}\label{I2x}
I_2^{TCSA}(x) = 2\cdot \left(\frac{2\pi}{R}\right)^2 \int_0^R dx\int_{-\infty}^0  dt \, \frac{|z|^{x}}{|z-1|^{2x}} .
\end{equation}
Expanding the integrands in powers of $z$ and $\bar z$, truncating with level $N_{tr}$, and performing
the integrals leaves us with (compare Eqn. (\ref{I2}))
\begin{eqnarray}\label{I2xsum}
I_2^{TCSA}(x) &=& 2\pi \sum^{N_{tr}-1}_{n=0} \frac{\Gamma^2(n+x)}{(n+\frac{x}{2})\Gamma^2(x)(n!)^2}\cr\cr
&\equiv& I^{TCSA}_{2,div.}(x)(N_{tr}(N_{tr}+x-1))^{x-1}+I^{TCSA}_{2,univ.}(x)+I^{TCSA}_{2,subleading}(x)N^{2x-6}_{tr}.
\end{eqnarray}
The coefficient of the first term in the above can be determined from the Euler-Maclaurin formula converting a sum to an integral
in combination with our exact result at $x=2$.  The result is
\begin{equation}
I^{TCSA}_{2,div.}(x) = \frac{2\pi}{\Gamma^2(x)(2x-2)}.
\end{equation}
The universal term, $I^{TCSA}_{2,univ.}$, can be determined from performing
the integral in Eqn. (\ref{I2x}) for $0 < x < 1$ and then analytically continuing \cite{Cardy}:
\begin{equation}
I^{TCSA}_{2,univ.}(x) = \frac{\pi\Gamma^2(\frac{x}{2})\Gamma(1-x)}{\Gamma^2(1-\frac{x}{2})\Gamma(x)}.
\end{equation}
Finally we determine $I^{TCSA}_{2,subleading}(x)$ numerically: we compute the sum in Eqn. (\ref{I2xsum}) numerically as a function of $N_{tr}$,
subtract the first two terms in the second line of Eqn. (\ref{I2xsum}), and fit the remainder to extract the coefficient
of $N_{tr}^{2x-6}$.  The results are plotted in Fig. (\ref{I2subleading}).  We note in particular that as $x \rightarrow 2$,
$I^{TCSA}_{2,subleading}(x) \rightarrow 0$.

\subsection{Evaluation of $I^{TCSA}_3(x)$}
$I^{TCSA}_3(x)$ is computed by first reexpressing $I_3(x)$ as an integral over the cylinder:
\begin{equation}
I_3(x) = (\frac{2\pi}{4}\bigg)^4\int^R_0 dx_1 dx_2\int^\infty_{-\infty} dt_1 dt_2 \frac{|z_1|^x|z_2|^x}{|z_1-z_2|^x|z_1-1|^x|z_2-1|^x},
\end{equation}
where $z_i=e^{2\pi (i x_i+t_i)/R}$.  Expanding the integrand in powers of $z_i$ and truncating these sums on the basis of a comparison
with the Lehmann expansion leaves us with
\begin{eqnarray}\label{I3sum}
I^{TCSA}_3(x) &=& 12\pi^2\sum^{N_{tr}-1}_{j=0}\sum^{N_{tr}-1}_{j=0}\sum^{j}_{\bar j=0}\sum^{N_{tr}-j-1}_{l=0}\sum^{N_{tr}-j-1}_{k=0}
\frac{\gamma_j\gamma_{\bar j}\gamma_l\gamma_{j+l-\bar{j}}\gamma_k\gamma_{k+j-\bar{j}}}{(j+l+\frac{x}{2})(j+k+\frac{x}{2})};\cr\cr
\gamma_j &=& \frac{\Gamma(j+\frac{x}{2})}{\Gamma(\frac{x}{2})\Gamma(j+1)}.
\end{eqnarray}
$I^{TCSA}_3(x)$ then has a similar structure to $I^{TCSA}_2(x)$:
\begin{equation}
I^{TCSA}_3(x) = I^{TCSA}_{3,div.}(x)(N_{tr}(N_{tr}+x-1))^{x-1}+I^{TCSA}_{3,univ.}(x)+I^{TCSA}_{3,subleading}(x)N^{x-2}_{tr}.
\end{equation}
We infer that the leading term of $I^{TCSA}_{3}(x)$ must be proportional to $(N_{tr}(N_{tr}+x-1))^{x-1}$ -- otherwise the
relationship between the TCSA cutoff, $N_{tr}$, and $\epsilon_p$ established in the evaluation of $I_2$ would breakdown.
(We will in any case verify this numerically in what is to come.)  The coefficient of the leading term can be found be converting the sums
to integrals:
\begin{equation}
I^{TCSA}_{3,div.}(x) = \frac{12\pi^2}{\Gamma^6(\frac{x}{2})}\int^1_0dj \int^j_0 d\bar{j} \int^{1-j}_0 dl \int^{1-j}_0 dk \frac{(j\bar{j}lk(l+j-\bar{j})(k+j-\bar{j}))^{\frac{x}{2}-1}}{(j+l)(k+l)}.
\end{equation}

\begin{figure}
\centering
\includegraphics[scale=0.9]{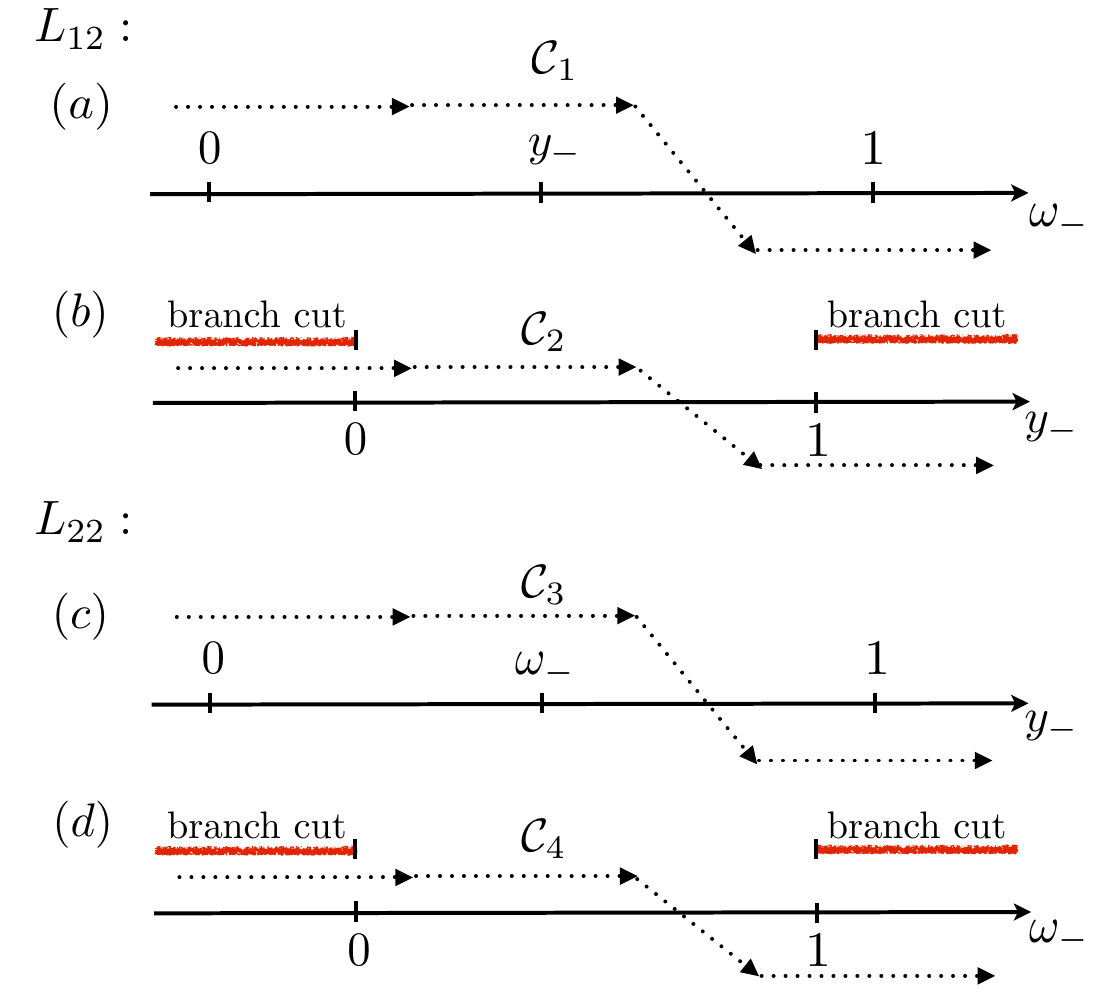}
\caption{Definition of contours used in evaluating
$I^{TCSA}_{3,univ.}(x)$.  There are branch points at all labelled points: $0,1,y_-,$ and $w_-$.  The red lines mark additional
branch cuts with branch points at $0$ and $1$ 
that result from integrating $w_-$ in the case of $L_{12}$ and $y_-$ in the case of $L_{22}$.}
\label{contours}
\end{figure}

We can also evaluate analytically $I^{TCSA}_{3,univ.}(x)$.  This can be computed by performing the integral in Eqn. (\ref{integrals})
for values of $x$ where it is convergent and then analytically continuing.  To make this evaluation we first
perform the integration by parts suggested in Ref. \cite{LudwigCardy}:
\begin{equation}
I^{TCSA}_{3,univ.}(x) = I_{3,univ.}(x) = \frac{2-x}{4-3x}\int d^2z_1 d^2z_2 \frac{|z_1+1|^{x-2}|z_2+1|^{x-2}}{|z_1-z_2|^x|z_1|^x|z_2|^x}\bigg(\frac{z_1}{1+z_1}+\frac{\bar{z_1}}{1+\bar{z_1}}\bigg).
\end{equation}
We now evaluate this integral following the techniques introduced in Ref. \cite{Dotsenko}
and used in Ref. \cite{sg_pt}.  Writing $z_i = x_i + iy_i$ and making the changes of variables:
\begin{equation}
y_i \rightarrow ie^{-i2\epsilon}y_i ,
\end{equation}
followed by
\begin{equation}
y_\pm = x_1 \pm y_1; ~~~~ w_\pm = x_2 \pm y_2 ,
\end{equation}
allows us to rewrite $I_3(x)$ as
\begin{eqnarray}
I_3(x) &=&\frac{2-x}{2(4-3x)}\int dy_+ (y_+-1-i\epsilon\Delta_y)^{\frac{x}{2}-1}(y_+-i\epsilon\Delta_y)^{-\frac{x}{2}}\cr\cr
&\times&\!\!\int dw_+ (w_+-y_+-i\epsilon(\Delta_w-\Delta_y))^{-\frac{x}{2}}(w_+-1-i\epsilon\Delta_w)^{\frac{x}{2}-1}(w_+-i\epsilon\Delta_w)^{-\frac{x}{2}}\cr\cr
&\times&
\!\!\int dy_- (y_--1+i\epsilon\Delta_y)^{\frac{x}{2}-2}(y_-+i\epsilon\Delta_y)^{-\frac{x}{2}+1}\cr\cr
&\times& \!\!\int dw_- (w_--y_-+i\epsilon(\Delta_w-\Delta_y))^{-\frac{x}{2}}(w_--1+i\epsilon\Delta_w)^{\frac{x}{2}-1}(w_-+i\epsilon\Delta_w)^{-\frac{x}{2}},
\end{eqnarray}
where $\Delta_{w/y}=(w/y)_+-(w/y)_-$.
The positions of the contours for $w_-$ and $y_-$ relative to the various branch cuts of the arguments depend
on the values of $w_+$ and $y_+$.  Only for certain values of $w_+$ and $y_+$ is it not possible to deform the
contours $w_-/y_-$ to infinity without encountering poles or branch cuts.  This allows us to restrict the limits
of $w_+$ and $y_+$ dramatically, simplifying the above
integral to
\begin{eqnarray}
I_3(x) &=& \frac{2-x}{2(4-3x)}(L_1+L_2);\cr\cr
L_1 &=& -L_{11}L_{12}; ~~~ L_2 = -L_{21}L_{22};\cr\cr
L_{11} &=& \int^1_0dw_+\int^{w_+}_0 dy_+(1-y_+)^{\frac{x}{2}-1}w_+^{-\frac{x}{2}}(w_+-y_+)^{-\frac{x}{2}}(1-w_+)^{\frac{x}{2}-1}y_+^{-\frac{x}{2}};\cr\cr
L_{12} &=&\int_{C_2}dy_- \int_{C_1} dw_- (1-y_-)^{\frac{x}{2}-2}y_-^{-\frac{x}{2}+1}(w_--y_-)^{-\frac{x}{2}}(1-w_-)^{\frac{x}{2}-1}w_-^{-\frac{x}{2}};\cr\cr
L_{21} &=& \int^1_0dy_+\int^{y_+}_0dw_+(1-y_+)^{\frac{x}{2}-1}w_+^{-\frac{x}{2}}(y_+-w_+)^{-\frac{x}{2}}(1-w_+)^{\frac{x}{2}-1}y_+^{-\frac{x}{2}};\cr\cr
L_{22} &=& \int_{C_3}dy_- \int_{C_4} dw_- (1-y_-)^{\frac{x}{2}-2}y_-^{-\frac{x}{2}+1}(y_--w_-)^{-\frac{x}{2}}(1-w_-)^{\frac{x}{2}-1}w_-^{-\frac{x}{2}}.
\end{eqnarray}
The contours $C_i,i=1,2,3,4$ are defined in Fig. (\ref{contours}).  These four separate integrals can now readily be
expressed in terms of generalized hypergeometric functions with the results found in Eqn. (\ref{evalI3}).

\begin{figure}
\centering
\includegraphics[scale=0.4]{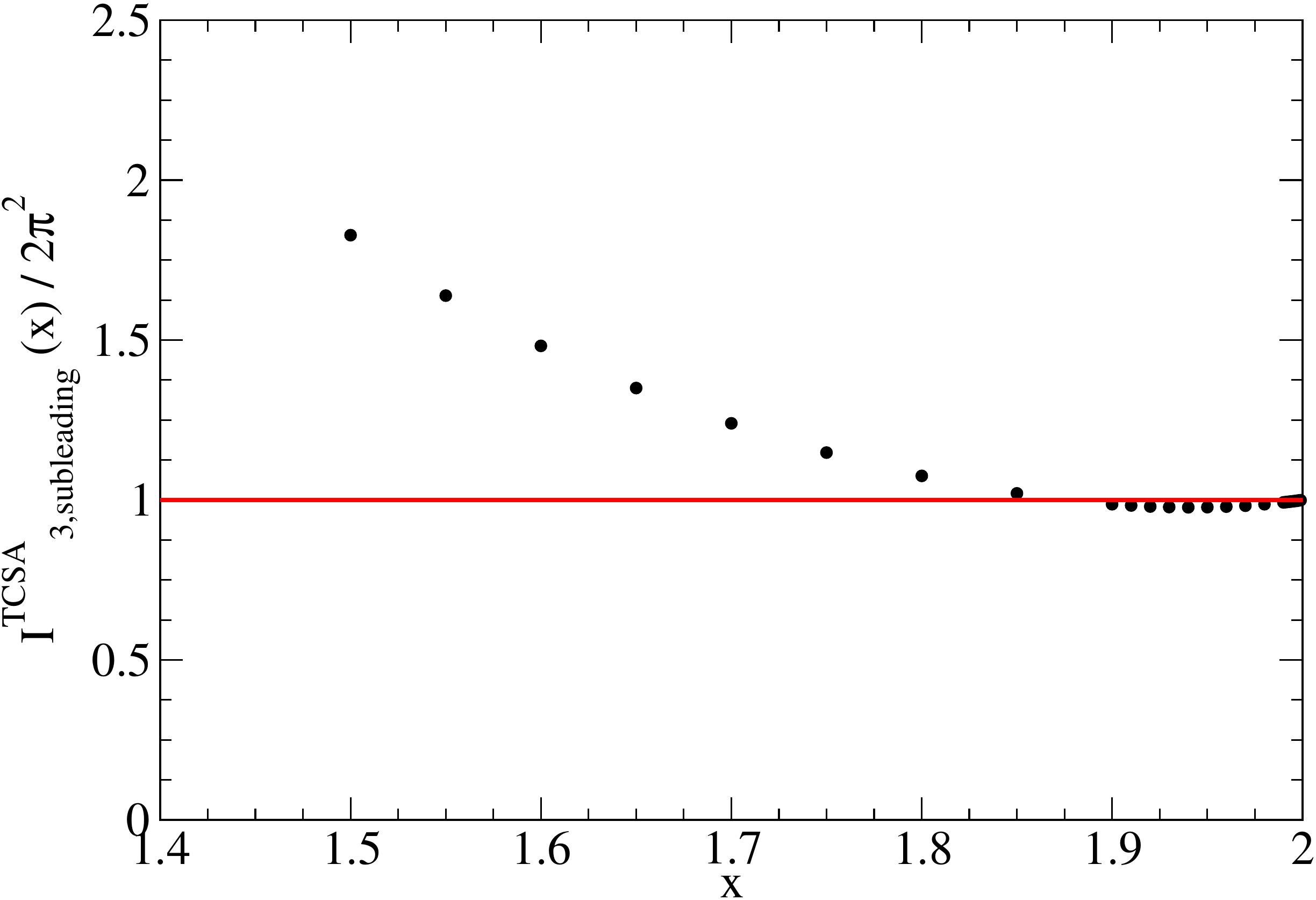}
\caption{Plot of  $I^{TCSA}_{3,subleading}(x)/(2\pi)^2$ as a function of $x$.}
\label{I3subleading}
\end{figure}

Finally we can evaluate $I^{TCSA}_{3,subleading}(x)$ numerically.  We compute this coefficient by evaluating the sum in Eqn. (\ref{I3sum})
for a range of $N_{tr}$,
subtracting off $I^{TCSA}_{3,div.}$ and $I^{TCSA}_{3,univ.}$.  The remainder is proportional to $N_{tr}^{x-2}$ and we
extract $I^{TCSA}_{3,subleading}$ as the fitting coefficient.  The results are found in Fig. \ref{I3subleading}.  We see in
particular that $I^{TCSA}_{3,subleading}(x) \rightarrow 2\pi^2$ as $x\rightarrow 2$.

\end{document}